\documentclass[11pt]{article}
\pdfoutput=1

\usepackage{a4wide}
\usepackage[fleqn]{amsmath}
\usepackage{hyperref}

\RequirePackage{amsmath,amssymb,amsxtra,amsthm}

\RequirePackage[T1]{fontenc}
\RequirePackage[utf8]{inputenc}

\RequirePackage{mathrsfs}
\RequirePackage{mathpazo}

\RequirePackage{setspace}
\RequirePackage{array}
\RequirePackage{booktabs}

\RequirePackage{braket}

\RequirePackage[pdftex,final]{graphicx}
\graphicspath{{Plots/}}

\usepackage{cite}

\newcommand{\be}{\begin{equation}}
\newcommand{\ee}{\end{equation}}
\newcommand{\bea}{\begin{eqnarray}}
\newcommand{\eea}{\end{eqnarray}}
\newcommand{\IR}{\mathbb{R}}
\newcommand{\IC}{\mathbb{C}}
\newcommand{\IZ}{\mathbb{Z}}

\newcommand{\IH}{\mathbb{H}}

\newcommand{\cF}{\mathcal{F}}

\newcommand{\cI}{\mathcal{I}}

\newcommand{\cM}{\mathcal{M}}
\newcommand{\cW}{\mathcal{W}}
\newcommand{\cN}{\mathcal{N}}

\newcommand{\cZ}{\mathcal{Z}}
\newcommand{\cO}{\mathcal{O}}

\newcommand{\de}{\mathrm{d}}

\newcommand{\I}{\mathrm{i}}

\newcommand{\sgn}{{\rm sgn}}

\renewcommand{\mod}{{\rm mod}}

\def\pa{\partial}

\numberwithin{equation}{section}
\numberwithin{table}{section}
\numberwithin{figure}{section}
\author{
  \begin{minipage}{0.97\linewidth}
    \vspace{1cm}
    \begin{center}
      \begin{small}
        \textbf{Carlo Angelantonj}$^{1}$, \textbf{Ioannis Florakis}$^{2}$ and  \textbf{Boris Pioline}$^{2,3}$
     \end{small}
    \end{center}
    \vspace{.3cm} \hspace{1.3cm}\begin{minipage}{.75\linewidth}
      {\it \begin{footnotesize}
          \begin{itemize}
          \item[${}^1$] Dipartimento di Fisica, Universit\`a di Torino, and INFN Sezione di Torino
          \\
            Via P. Giuria 1, 10125 Torino, Italy
         \item[${}^2$] CERN Dep PH-TH, 1211 Geneva 23, Switzerland
         \item[${}^3$] Laboratoire de Physique Th\'eorique et Hautes Energies, CNRS UMR 7589,
         \\
         Universit\'e Pierre et Marie Curie - Paris 6, 4 place Jussieu,
         75252 Paris cedex 05, France
          \end{itemize}
        \end{footnotesize}}
    \end{minipage}
    \vspace{1cm}
  \end{minipage}
}

\date{}

\title{\vspace{3cm}
  \begin{huge}
    \textbf{Threshold corrections,\\ generalised prepotentials \\ and Eichler integrals}
  \end{huge}
}

\begin{document}

\begin{titlepage}
  \maketitle
  \thispagestyle{empty}

  \vspace{-14cm}
  \begin{flushright}
    CERN-PH-TH/2015-011 
   \end{flushright}

  \vspace{11cm}

\begin{center}
\textsc{Abstract}
\\
\end{center}
We continue our study of one-loop integrals associated to BPS-saturated amplitudes in $\cN=2$ heterotic vacua. We compute their large-volume behaviour, and express them as  Fourier series in the complexified volume, with Fourier coefficients given in terms of Niebur-Poincar\'e series in the complex structure modulus. The closure of Niebur-Poincar\'e series under modular derivatives  implies that such integrals derive from holomorphic prepotentials $f_n$, generalising the familiar prepotential of $\cN =2$ supergravity. These holomorphic prepotentials transform anomalously under T-duality, in a way characteristic of Eichler integrals. We use this observation to compute their quantum monodromies under the duality group. We extend the analysis to modular integrals with respect to Hecke congruence subgroups, which naturally arise in compactifications on non-factorisable tori and freely-acting orbifolds. In this case, we derive new explicit results  including closed-form expressions for  integrals involving the  $\varGamma_0 (N)$ Hauptmodul, a full characterisation of holomorphic prepotentials including their quantum monodromies, as well as concrete formul\ae\ for holomorphic Yukawa couplings. 
\vfill
{\small
\begin{itemize}
\item[E-mail:] {\tt carlo.angelantonj@unito.it}\\ {\tt ioannis.florakis@cern.ch}\\
{\tt boris.pioline@cern.ch}
\end{itemize}
}
\vfill

\end{titlepage}

\setstretch{1.1}

\tableofcontents


\section{Introduction}

The leading radiative corrections to scattering processes in closed string theories compactified on a $d$-dimensional torus are computed by modular integrals of the form
\begin{equation}
\cI = \int_\cF d\mu \, \varGamma_{d+k,d}  \, \varPhi  \,,
\label{intintro}
\end{equation}
where $\varGamma_{d+k,d}$ is the partition function of the Narain  lattice with signature $(d+k,d)$, 
$\varPhi$ is an appropriate modular form of negative 
weight $(-\frac{k}{2},0)$ under the modular group
$\varGamma={\rm SL}(2;\IZ)$ of the worldsheet torus, 
and $\cF$ is a fundamental domain for the action
of $\varGamma$ on the Poincar\'e upper half plane $\IH$. In order to extract quantitative predictions from string theory, it is therefore essential to develop efficient tools to compute such modular integrals. 

In the traditional lattice unfolding method  \cite{O'Brien:1987pn,McClain:1986id,Dixon:1990pc}, one decomposes the lattice sum $\varGamma_{d+k,d}$ into orbits under the modular group, and reduces each orbit to a single term while unfolding the integration domain. Although the resulting expression is very useful in extracting the asymptotic expansion of $\cI$ in the limit where the volume of the torus $T^d$ becomes large, it depends on a choice of chamber in the Narain moduli space,  and T-duality invariance is no longer manifest.

In a series of recent works \cite{Angelantonj:2011br,Angelantonj:2012gw,Angelantonj:2013eja}, we proposed a new method, which assumes that  $\varPhi$ can be written as an absolutely convergent Poincar\'e series under the modular group. Upon unfolding the integration domain against the sum over images,  and evaluating the integral, the result is expressed as a sum over charges satisfying a BPS-type condition. This can be recognised as a particular type of Poincar\'e series for the automorphism group $ {\rm O} (d+k,d;\mathbb{Z} )$ of the Narain lattice, which is thus manifestly invariant under T-duality. The analytic structure of the integral near points of enhanced gauge symmetry is also manifest, while the large volume asymptotics is no longer apparent. In fact, the method can be applied even when the integrand does not include the Narain lattice, {\it e.g.} in ten-dimensional string vacua.

If $\varPhi$ is a generic non-holomorphic modular form, there is no general known way to represent it as a Poincar\'e series. For  BPS-saturated amplitudes, however, the modular form $\varPhi$ turns out to be  weakly almost holomorphic, {\it i.e.} a polynomial in the ring $\IC[\hat E_2,E_4,E_6,1/\Delta]$. In that case, it can actually be decomposed as a linear combination of so-called Niebur-Poincar\'e series $\cF(s,\kappa,w;z)$  \cite{1004.11021, Angelantonj:2012gw}. These Poincar\'e series, introduced in the mathematics literature in \cite{0288.10010,0543.10020},  have the virtues of  being eigenmodes of the Laplacian on $\IH$ with a pole of order\footnote{While the integrands appearing in  heterotic string theory only have a first order pole at $z=\infty$, it is useful to allow poles of arbitrary order $\kappa\geq 1$ as such poles typically occur in modular integrals for congruent subgroups, relevant for orbifold compactifications. Moreover, Niebur-Poincar\'e series $\cF(s,\kappa',w)$ with $\kappa'>1$ occur in the Fourier expansion of modular integrals of the form \eqref{intintros}, even if $\kappa=1$.} $\kappa$ at the cusp $z=\infty$ and of being closed under modular derivatives and Hecke operators. In addition, they reduce to harmonic Maass forms when $s=1-\tfrac{w}{2}$, or to modular derivatives thereof when $s=1-\tfrac{w}{2}+n$ with $n$ a positive integer. 
For suitable linear combinations, the non-analytic parts  of $\cF(s,\kappa,w;z)$ (encoded in the `shadow') cancel, producing the desired weakly almost holomorphic form $\varPhi$.
The  integral \eqref{intintro}
is thus reduced to a sum of integrals of the form
\begin{equation}
\cI (s,\kappa)= \int_\cF d\mu \, \varGamma_{d+k,d}  \, \cF(s,\kappa,-k/2)  \,,
\label{intintros}
\end{equation}
each of which can be computed by unfolding the fundamental domain against the Niebur-Poincar\'e series \cite{Angelantonj:2012gw}. 

In this work, we compute the large volume expansion of the modular integral \eqref{intintros},
thus providing an alternative way to extract the asymptotics of  \eqref{intintro}
within our unfolding method. The advantage of decomposing  $\varPhi$ into a sum of 
Niebur-Poincar\'e series is that  the integral \eqref{intintro} is then decomposed 
into a sum of terms which are separately eigenmodes of the Laplacian on 
the Narain moduli space
\begin{equation}
\label{Narainmod}
G_{d+k,d} = {\rm O} (d+k,d;\mathbb{Z} ) \backslash {\rm SO} (d+k,d ) / {\rm SO} (d+k ) \times {\rm SO} (d) \, ,
\end{equation}
and which behave in a simple way under covariant derivatives with respect to the moduli.

For simplicity, we  restrict our attention to integrals involving lattices of signature (2,2), where the moduli space \eqref{Narainmod} is parametrised by the standard variables $(T=T_1+\I T_2,U=U_1+\I U_2)\in \IH_T \times \IH_U$. They correspond to the K\"ahler modulus and complex structure of a two-torus, and are identified under separate actions of the modular group ${\rm SL}(2;\IZ)$ on $T$ and $U$, and under the involution  $\sigma:T\leftrightarrow U$.
This case is relevant when the internal space includes a $T^2$ factor, {\it e.g.} in the context of heterotic strings compactified on $T^2$ (with 16 supercharges),  on ${\rm K3}\times T^2$ (with 8 supercharges), or on $\cN =1$ orbifolds with $\cN =2$ sub-sectors, upon restriction to backgrounds with trivial or discrete Wilson lines\footnote{We defer the analysis of asymmetric lattices with non-trivial Wilson lines to a forthcoming publication.}. We consider both the limit where the volume $T_2$  is scaled to infinity keeping the complex structure modulus $U$ fixed, and the limit where both $T_2$ and $U_2$ are scaled to infinity (in which case the asymptotic expansion depends on a choice of chamber, $T_2>U_2$ {\it vs.} $T_2<U_2$). In the latter case, the result is expressed as an infinite sum of a special type of single-valued polylogarithms, known in the mathematics literature as Bloch-Wigner-Ramakrishnan polylogarithms  \cite{zbMATH04144378}, which coincide with the real part of the `combined polylogarithms' introduced in \cite{Kiritsis:1997hf}. 
In the former case, $T_2\to \infty$ with $U$ fixed, we show that the Fourier coefficients of the integral $\cI(s,\kappa)$ with respect to $T_1$ are given by the same 
Niebur-Poincar\'e series $\cF(s,\kappa,0)$ present in the integrand, now as functions of the $U$ modulus. This is a manifestation of a phenomenon first observed in \cite{Bachas:1997mc} which, as we shall see, can be traced to the average value property of eigenmodes of the Laplacian on the Poincar\'e upper half-plane.  An important consequence of this observation is that, using the closure of Niebur-Poincar\'e series under modular derivatives,  one may express the integral \eqref{intintros}   as an iterated derivative of a `harmonic prepotential' $h_n$\cite{Kiritsis:1997hf}
\be
\label{iderpot0}
{\mathcal I} (n+1,\kappa) = 4 \, {\rm Re}\, \left[  \frac{(-D_T  D_U)^n}{n!} h_n(T,U) \right]
- 24\, \sigma_1(\kappa)\, \delta_{n,0} \, \log T_2 U_2\ .
\ee
The  prepotential  $h_n$ is by construction harmonic in $(T,U)$  and transforms covariantly under ${\rm SL}(2;\IZ)_T\times {\rm SL}(2;\IZ)_U$, while the last term originates from the infrared divergence of the integral when $n=0$. Crucially, it is also possible to construct a {\it holomorphic} prepotential 
$f_n(T,U)$  which satisfies the same relation \eqref{iderpot0}.  This holomorphic prepotential is
well-defined up to the addition of polynomials of degree $2n$ in $T$ and $U$ with real coefficients, which lie in the kernel of the operator ${\rm Re}(D_T  D_U)^n$. As a result, it need not transform homogeneously under T-duality, but may pick up an additional degree $2n$ polynomial. This type of anomalous behaviour under modular transformations is characteristic of Eichler integrals. Indeed, we trace the modular anomaly of $f_n(T,U)$ to the presence of a term proportional to the Eichler integral of the holomorphic Eisenstein series $E_{2n+2}$ in the Fourier expansion. We use this fact to fully characterise the T-duality transformations of the generalised prepotential $f_n$. 

From a physics viewpoint, the prepotentials $f_n$ for $n>1$ can be viewed as generalisations of the usual holomorphic prepotential $F$ that governs $\cN=2$ supergravity in four dimensions -- or 
more accurately, of the one-loop contribution to it \cite{Antoniadis:1995ct,deWit:1995zg,Harvey:1995fq,Forger:1997tu}. In fact, it was already recognised in  \cite{Antoniadis:1995ct} that the quadratic terms  appearing in the behaviour of the one-loop prepotential $f_1$ under monodromies around singularities in moduli space can be obtained from contour integrals of suitable modular forms, which are Eichler integrals in essence. Our results enable us  to recover the monodromy matrices computed in  \cite{Antoniadis:1995ct} in a streamlined fashion, and to  extend the analysis to $n>1$. The case $n=2$ was studied in the context of $F^4$ couplings in heterotic strings compactified on $T^2$ \cite{Lerche:1998nx,Lerche:1999hg}, where a relation of the generalised prepotentials 
$f_n$ to periods on symmetric products of $n$ K3's was suggested. However, the analogue of special geometry for $n>1$ is not understood at present. As a step in this direction, we construct
a class of `generalised Yukawa couplings', obtained from $f_n$ by iterated derivatives with respect to the $T$ and $U$ moduli, which compute modular integrals with insertions of momenta and are free of modular anomalies.

From a mathematics viewpoint, the prepotentials $f_n$ can also be viewed as interesting generalisations of Borcherds automorphic products \cite{zbMATH00220742} -- or rather, logarithms thereof.  Indeed, in the special case where $\cF(1+n,1,-2n)$
is a weakly holomorphic modular form, $f_n$ can be written schematically as a polylogarithm sum
\be
f_n(T,U)= 
\frac{1}{2^{2n}} 
\sum_{N,M} F_n (NM) \, {\rm Li}_{2n+1} (q_T^M\, q_U^N) + P_{2n}(T,U)\ ,
\ee
where $F_n(M)$ are the Fourier coefficients of 
$\cF(1+n,1,-2n)$ and $P_{2n}$ is the polynomial ambiguity mentioned above. For $n=0$, ${\rm Li}_1(z)=-\log(1-z)$, so $f_0$ is recognised as the logarithm of a
Borcherds automorphic product. For $n>0$, $f_n$ is no longer the logarithm of a product, but it provides an interesting example of an holomorphic `Mock' modular form on $(\IH_T\times \IH_U)/\sigma$ with known singular behaviour at all divisors.

Returning to physics, the integrals \eqref{intintro} do not exhaust the class of one-loop amplitudes of interest in string theory. In fact, in compactifications on non-factorisable tori \cite{Mayr:1993mq,Bailin:2014nna} or on freely-acting orbifolds \cite{Kiritsis:1998en, Kiritsis:1997ca, Angelantonj:2006ut, Angelantonj:2014dia, Faraggi:2014eoa} the integrand is typically expressed as a sum of various sectors $(h,g)$ involving shifted Narain lattices $\varGamma_{2,2}[{ h \atop g}]$ coupled to modular forms $\varPhi[{ h \atop g}]$.
Although the full integrand is, by construction, invariant under the full modular group ${\rm SL} (2;\mathbb{Z})$, each term is only invariant under a suitable congruence subgroup. These setups have not been extensively analysed so far in the literature and call for a systematic investigation. 
Upon partial unfolding \cite{Angelantonj:2013eja}, one may compute the integral by focusing on a single sector now integrated over the fundamental domain of the congruence subgroup.  In the case of freely-acting $\mathbb{Z}_N$ orbifolds of $\cN =2$ compactifications, the relevant  subgroup is the Hecke congruence subgroup $\varGamma_0 (N)$. In parallel
with  the ${\rm SL} (2;\mathbb{Z})$ case, one may decompose $\varPhi[{ h \atop g}]$ 
as a linear combination of the $\varGamma_0 (N)$ Niebur-Poincar\'e series $\cF_\mathfrak{a} (1+n,\kappa , 0)$ attached to the cusp $\mathfrak{a}$ \cite{Angelantonj:2013eja}. One may then evaluate the modular integral by  unfolding the fundamental domain $\cF_N$ of $\varGamma_0 (N)$ against the Poincar\'e series. This procedure was suggested in \cite{Angelantonj:2013eja} and applied to the study of radiative corrections to gauge couplings in classically stable non-supersymmetric heterotic vacua in \cite{Angelantonj:2014dia}. The result is a representation of the modular integral
as a Poincar\'e series under the T-duality group,  in general a congruence subgroup
of ${\rm O}(2,2;\IZ)$.

In this work, we continue this analysis and extract the large-volume expansion of one-loop modular integrals for $\varGamma_0(N)$ with $N$ prime. We obtain expressions which extend the results for ${\rm SL}(2;\mathbb{Z})$ and are nicely  covariant with respect to the choice of cusp $\mathfrak{a}$. The advantage of our approach, based on the special properties of $\cF_\mathfrak{a} (1+n, \kappa , 0)$, is that we  obtain a $T$-Fourier series expansion with $U$-dependent coefficients which are manifestly invariant under $\varGamma_0 (N)_U$, and which actually are given in terms of the same class of Niebur-Poincar\'e series. This property is instrumental in allowing us to derive new explicit results. Namely, in the $n=0$ case we re-sum the integral in terms of the $\varGamma_0 (N)$ Hauptmodul by generalising the celebrated ${\rm SL} (2;\mathbb{Z})$  Borcherds' product formula to the case of Hecke congruence subgroups. For $n\ge 1$ we identify  generalised holomorphic prepotentials and fully characterise their transformation properties under T-duality. Finally, we express the generalised holomorphic Yukawa couplings in terms of the standard generators of the ring of $\varGamma_0 (N)$ modular forms.

The paper is organised as follows: in Section \ref{sec_intsl2z}, after recalling basic properties of Niebur-Poincar\'e and Eisenstein series, we compute the large volume asymptotics of the modular integral \eqref{intintros} or, equivalently, the Fourier expansion with respect to $T_1$, starting from its T-duality invariant representation. We express the results in terms of single-valued polylogarithms, and make contact with the literature. In Section 
\ref{sec_prepot} we integrate this result in terms of generalised harmonic and holomorphic prepotentials, explain their relation to Eichler integrals, and determine their anomalous transformation properties under T-duality. We construct generalised Yukawa couplings by taking suitable modular
derivatives of the generalised prepotentials, and connect them to modular integrals with lattice-momentum insertions. Section \ref{HeckeIntegrals_sec} extends the previous analysis to modular integrals involving shifted 
Narain lattices. We obtain new explicit results including  `cusp-covariant' Fourier series expansions, closed expressions for special classes of integrals involving the  $\varGamma_0 (N)$ Hauptmodul,
a full characterisation of the generalised holomorphic prepotentials, as well as concrete formul\ae\ for the generalised holomorphic Yukawa couplings. Appendix A collects our conventions for the Kloosterman sums and zeta functions, and Appendix B provides a detailed derivation of the modular properties of the relevant Eichler integrals.

\section{Modular integrals for ${\rm SL}(2;\IZ)$ \label{sec_intsl2z}}

In this section, we shall focus on modular integrals \eqref{intintro} involving the product of a  signature (2,2) Narain lattice $\varGamma_{2,2}$ times a weakly almost holomorphic modular functions $\varPhi (\tau )$. As anticipated, the latter can be uniquely decomposed in terms of Niebur-Poincar\'e series of zero weight \cite{1004.11021, Angelantonj:2012gw}. Thus, we begin by collecting some useful properties of Eisenstein and Niebur-Poincar\'e series which are instrumental for the subsequent analysis.

\subsection{Niebur-Poincar\'e series for ${\rm SL}(2; \IZ)$}\label{secNPS}

The Niebur-Poincar\'e series \cite{ 0288.10010, 0543.10020,1004.11021}\footnote{We shall henceforth omit the explicit dependence of the Niebur-Poincar\'e series on the modular parameter $z$ whenever it is clear from the context.} 
\begin{equation}
\begin{split}
\label{Fskw}
\cF(s,\kappa,w ; z) =&\tfrac12 \sum_{\gamma\in \varGamma_\infty\backslash \varGamma} \,
\cM_{s,w}(-\kappa y)\, e^{-2\pi\I\kappa x}\, \vert_w\, \gamma 
\\
=& \tfrac{1}{2} \sum_{(c,d)=1} (cz + d)^{-w}\, {\mathcal M}_{s,w} \left(- \frac{\kappa\, y}{|cz + d|^2} \right) \, \exp \left\{ -2\I\pi \kappa \left( \frac{a}{c} -\frac{cx+d}{c|cz + d|^2}\right)\right\} \, .
\end{split}
\end{equation}
defines a one-parameter family of modular forms of non-positive weight $w$ and order $\kappa$ pole in the nome $q=e^{2i\pi z}$ at the cusp $z\equiv x+\I y =\infty$. Its seed is expressed in terms of the Whittaker 
$M$-function
\begin{equation}
\cM_{s,w}(t) = |4\pi t|^{-\frac{w}{2}}\, M_{\frac{w}{2}\sgn(t), s-\frac12}  \left( 4\pi | t| \right) \,,
\label{curlyM}
\end{equation}
whose behaviour at $t\to 0$ guarantees absolute convergence when ${\rm Re}\, (s)>1$. 
The Poincar\'e series can be shown to admit a meromorphic continuation to the complex $s$ plane.
Moreover, in the domain of analyticity, $\cF (s,\kappa , w) $   is an eigenmode of the weight-$w$ hyperbolic Laplacian $\varDelta_{z}$,
\begin{equation}
\label{laplEskw}
\left[\varDelta_z + \tfrac{1}{2}\,  s(1-s) +\tfrac{1}{8}\, w(w+2)\right] \, \cF(s,\kappa,w) = 0\ ,
\end{equation}
where
\begin{equation}
\varDelta_{z} = 2\, \bar D_z \, D_{z}\,, \qquad {\rm with} \quad \bar D_z = -i \pi \, y^2 \partial_{\bar z} \,, \quad D_{z} =  \frac{i}{\pi} \left( \partial_z - \frac{iw}{2y}\right)\,.
\end{equation}
Its Fourier expansion reads
\begin{equation}
\label{FskwF}
\cF(s,\kappa,w)=\cM_{s,w}(-\kappa y)\, e^{-2i \pi \kappa x}
+ \sum_{m\in\IZ} \, \tilde\cF_m(s,\kappa,w; y) \, e^{2 i \pi  m x}\,,
\end{equation}
where, in the mode with frequency $m=-\kappa$, we have separated  the contribution of the seed. Explicitly, 
\begin{equation}
\label{NPFouriermodes}
\begin{split}
\tilde \cF_0 (s,\kappa , w ; y) &= f_0 (s,\kappa , w) \, y^{1-s-\frac{w}{2}}\,,
\\
\tilde \cF_m (s, \kappa , w ; y ) &= f_m (s,\kappa , w)\, \cW_{s,w} (my)\,,
\end{split}
\end{equation}
with
\begin{equation}
\label{NPFouriermodes2}
\begin{split}
f_0 (s,\kappa,w) &= 
\frac{2^{2-w}\, i^{-w}\, \pi^{1+s-\frac{w}{2}}\, |\kappa|^{s-\frac{w}{2}}\, \varGamma (2s-1)\, \sigma_{1-2s} (\kappa )}{\varGamma(s-\frac{w}{2}) \, \varGamma(s+\frac{w}{2})\, \zeta(2s)} \,,
\\
f_m (s,\kappa , w) &= \frac{4\pi\,i^{-w}\, |\kappa| \, \varGamma(2s)}{ \varGamma ( s+\frac{w}{2} \, {\rm sgn}(m))}\,  \left|  \frac{m}{\kappa}\right|^{\frac{w}{2}} \,  \cZ (m,-\kappa ; s) \,.
\end{split}
\end{equation}
In these expressions, $\cW_{s,w} (t)$ is related to the Whittaker $W$-function
\begin{equation}
\cW_{s,w} (t) = |4\pi t |^{-\frac{w}{2}}\, W_{\frac{w}{2}\sgn(t), s-\frac12}
\left(4\pi |t| \right) \,,
\end{equation}
$\sigma_\ell (n)$ is the divisor function, while $\cZ (a,b; s)$ is the {\em associated} Kloosterman--Selberg zeta function defined in \eqref{KSzetadef}.

Niebur-Poincar\'e series are closed under the action of the modular derivatives,
\begin{equation}
\begin{split}
D_z\, \cF (s,\kappa , w) &= 2 \kappa (s+\tfrac{w}{2}) \, \cF (s,\kappa , w+2 )\,,
\\
\bar D_z \, \cF (s,\kappa ,w) &= \frac{1}{8\kappa} (s- \tfrac{w}{2}) \, \cF (s,\kappa , w-2)\,,
\end{split}
\label{NPderiv}
\end{equation}
and of the Hecke operator\footnote{The Hecke operator $H_n^{(z)}$ acts on the Fourier modes of a weight-$w$ modular form $\varPhi = \sum_m  \varPhi (m,y) \,e^{2i \pi m x}$ as
\begin{equation}
H_n^{(z)}  \, \varPhi (m,y) = n^{1-w} \sum_{d|(m,n)} d^{w-1}\, \varPhi (nm/d^2 , d^2 y/n)\,. \nonumber
\end{equation}
It satisfies the commutative
algebra $H_\kappa^{(z)}  \, H_\lambda^{(z)}  = \sum_{d|(\kappa , \lambda)} d^{1-w}\, H^{(z)} _{\kappa \lambda / d^2}$.}
\begin{equation}
\cF (s,\kappa , w) = H_\kappa^{(z)} \, \cF(s, 1 ,w)\, .
\label{NPHecke}
\end{equation}

For the special values $s = \frac{w}{2}$ and $s=1-\frac{w}{2}$, within the domain of analyticity, 
$\cF (s,\kappa ,w)$ becomes a harmonic Maa\ss\ form. For negative weight, the case of interest for most applications to string theory \cite{Angelantonj:2012gw}, the Niebur-Poincar\'e series
is absolutely convergent at the latter value $s=1-\frac{w}{2}$.
Moreover, since the first cusp form for ${\rm SL} (2;\IZ)$ appears at weight twelve, for $w \in \{ 0,-2,-4,-6,-8,-12\}$ the shadow of $\cF(1-\frac{w}{2},\kappa,w)$ vanishes and the Niebur-Poincar\'e series is actually holomorphic, hence valued in the polynomial ring generated by $E_4$, $E_6$ and $\varDelta^{-1}$, where $E_{2n}$ are the holomorphic Eisenstein series and $\varDelta$ is the modular discriminant. Other Niebur-Poincar\'e series of interest are $\cF(1-\frac{w}{2} +n,\kappa,w)$, with $n$ a positive integer, which can be reached from the harmonic Maa\ss\ form 
$\cF(1-\frac{w}{2} +n,\kappa,w-2n)$ by acting with $n$ modular derivatives. In fact, one can prove \cite{1004.11021, Angelantonj:2012gw, Angelantonj:2013eja} that any weak almost holomorphic modular form of negative weight in the polynomial ring generated by $\hat E_2 $, $E_4$, $E_6$ and $\varDelta^{-1}$, with $\hat E_2$ being the almost holomorphic Eisenstein series of weight two, can be uniquely decomposed into a linear combination of Niebur-Poincar\'e series,
\begin{equation}
\varPhi (z) = \sum_{0<m\le\kappa}\, \sum_{\ell =0}^n \, d_\ell (m) \, \cF \left(1-\tfrac{w}{2}+\ell , m , w \right)\,.
\label{lineardec}
\end{equation}
The coefficients $d_\ell (m)$ are determined by the principal part of the $q$-Laurent expansion of $\varPhi$ itself \cite{Angelantonj:2013eja}.
The integer $n$ counts the maximal power of $\hat E_2$ in the ring decomposition of $\varPhi$, and is known as the {\em depth}. An important fact is that for each value of $\ell$, the shadows cancel in the linear combination \eqref{lineardec} . 

It will also be useful to introduce the non-holomorphic weight-$w$ Eisenstein series
\begin{equation}
\begin{split}
E(s,w; z) &= \tfrac{1}{2} \sum_{\gamma\in\varGamma_{\infty}\backslash\varGamma} \,  y^{s-\frac{w}{2}}\vert_w\, \gamma 
\\
&= \tfrac{1}{2} \sum_{(c,d)=1} \frac{y^{s-\frac{w}{2}}} {(c\, z+d)^{w} \, |c\, z+d|^{2s-w}} 
\end{split}
\label{defEsl2}
\end{equation}
which is absolutely convergent for ${\rm Re}\, (s)>1$, and admits the Fourier series expansion
\be
\label{FourierEisw}
\begin{split}
E(s,w;z)=& y^{s-\frac{w}{2}}+
\frac{4^{1-s}\, i^{-w} \pi \varGamma(2s-1) \zeta(2s-1)}{\varGamma(s+\frac{w}{2})\, \varGamma(s-\frac{w}{2})\,\zeta(2s)} y^{1-s-\frac{w}{2}}
\\
&+ \sum_{m\neq 0}
\frac{i^{-w} (4\pi |m|)^{\frac{w}{2}} \pi^s}{\varGamma(s+\tfrac{w}{2} {\rm sgn} (m))\, \zeta(2s)}\,
|m|^{-s}\, \sigma_{2s-1}(|m|)\, \cW_{s,w}(m y)\, e^{2i \pi mx}\,.
\end{split}
\ee
The Eisenstein series $E(s,w)$ can be obtained formally as the limit $\kappa \to 0$ of $|4\pi \kappa |^{\frac{w}{2}-s}\, \cF (s,\kappa , w) $, and thus shares several of the properties of the Niebur-Poincar\'e series. Namely, it is an eigenmode of the weight-$w$ hyperbolic Laplacian 
\begin{equation}
\left[ \varDelta_w+\tfrac12 s(1-s)+\tfrac18 w(w+2) \right] \, E(s,w;z) =0\, ,
\end{equation}
is closed under the action of the modular derivatives 
\begin{equation}
\begin{split}
D_z\, E(s,w;z) &= \frac{1}{2\pi} (s+\tfrac{w}{2})\, E(s,w+2;z)\,,
\\
\bar D_z\, E(s,w;z) &= \frac{\pi}{2} (s-\tfrac{w}{2})\, E(s,w-2;z)\,,
\end{split}
\label{EPderiv}
\end{equation}
and becomes harmonic in $z$ when $s=\tfrac{w}{2}$ or $s=1-\tfrac{w}{2}$. For $w>2$, $E (\tfrac{w}{2}, w)$ coincides with the  holomorphic Eisenstein series $E_w $.

\subsection{Evaluation of the modular integral}

We can now proceed to the study of the integral \eqref{intintro}. Since any weak almost holomorphic modular form $\varPhi$ can be decomposed as in \eqref{lineardec} it is sufficient to focus our attention on the basic regularised modular integral
\begin{equation}
\cI (s,\kappa ;T,U) = {\rm R.N.} \int_\cF d\mu \, \varGamma_{2,2} (T,U; \tau ) \, \cF (s,\kappa , 0; \tau)\,,
\label{modint}
\end{equation}
where $\varGamma_{2,2} (T,U;\tau)$ is the partition function of the even self-dual lattice of signature $(2,2)$, and ${\rm R.N.}$ stands for the modular invariant renormalisation prescription of 
\cite{MR656029, Angelantonj:2011br,  Angelantonj:2012gw}
for treating the infra-red divergences arising at the cusp. For $s=1$, the  integral
\eqref{modint} develops a simple pole, which must be subtracted  \cite{1004.11021, Angelantonj:2012gw}. By abuse of notation we denote by $\cI(1,\kappa;T,U)$ the result of this subtraction. 
The partition function 
\begin{equation}
\label{G22ham}
\varGamma_{2,2} (T,U;\tau) = 
\tau_2\, \sum_{m_1,m_2,n^1,n^2\in\IZ} q^{\frac14 |p_L|^2} \bar q^{\frac14 |p_R|^2} \,,
\end{equation}
which we refer to as the Narain lattice, is a sum over  integers $m_i$ and $n^i$ identified with the Kaluza-Klein momentum and winding numbers along a two-dimensional torus. The left-moving and right-moving momenta on the two-dimensional lattice
\begin{equation}
\begin{split}
p_{\rm R} &= \frac{m_2 - U\, m_1 + T n^1 + T U\, n^2 }{\sqrt{T_2\, U_2}}  \,,
\\
p_{\rm L} &= \frac{m_2 - U\, m_1 +\bar  T  n^1 + \bar T U\, n^2  }{\sqrt{T_2\, U_2}}  \,,
\end{split}
\end{equation}
depend on the K\"ahler modulus $T=T_1 + i T_2$ and the complex structure modulus $U=U_1 + i U_2$ of the two-torus. 
$\varGamma_{2,2} (T,U;\tau)$ is modular invariant under ${\rm SL} (2;\IZ)$, which
requires the modular weight of the Niebur-Poincar\'e series in \eqref{modint} to vanish.

Eq. \eqref{G22ham} corresponds to the so-called Hamiltonian representation of the Narain lattice, and  exhibits manifest invariance under the duality group ${\rm O} (2,2;\mathbb{Z} ) \sim {\rm SL} (2;\mathbb{Z})_T \times {\rm SL} (2;\mathbb{Z} )_U \ltimes \sigma_{T,U}$, where the $\mathbb{Z}_2$ generator $\sigma_{T,U}$ exchanges the $T$ and $U$ moduli.
A Poisson summation over the momenta $m_i$ yields the so-called Lagrangian representation
\begin{equation}
\label{G22lag}
\varGamma_{2,2}(T,U;\tau)=
T_2\, \sum_{A\in {\rm Mat}_{2\times 2} (\mathbb{Z})}  e^{-2\pi\I T\,\det (A)}
\exp\left[ -\frac{\pi T_2}{\tau_2 U_2} \left| \begin{pmatrix}1 & U \end{pmatrix} \, A \, \begin{pmatrix}- \tau \\ 1 \end{pmatrix} \right|^2 \right]\,,
\end{equation}
which now exhibits manifest invariance under ${\rm SL} (2;\mathbb{Z})_\tau \times {\rm SL} (2;\mathbb{Z})_U \ltimes \sigma_{\tau , U}$. As a consequence, $\varGamma_{2,2}(T,U;\tau)$ is invariant under the action of ${\rm SL} (2;\mathbb{Z})_\tau \times {\rm SL} (2;\mathbb{Z})_T \times {\rm SL} (2;\mathbb{Z})_U \ltimes \sigma_{\tau , T, U}$, where $\sigma_{\tau , T, U}$ permutes the $\tau$, $T$ and $U$  moduli. This triality results in the identities 
\be
\label{DelTUtau}
\varDelta_T\, \varGamma_{2,2} \equiv \varDelta_U \varGamma_{2,2} \equiv \varDelta_{\tau} \varGamma_{2,2}\ ,\qquad
H_n^{(\tau)} \varGamma_{2,2}\equiv
H_n^{(T)} \varGamma_{2,2} \equiv
H_n^{(U)} \varGamma_{2,2} \, ,
\ee
where $\varDelta_z$ is the weight-zero hyperbolic Laplacian in the variable $z$. Together with the differential equation \eqref{laplEskw}, the first equation in \eqref{DelTUtau} implies that the modular integral \eqref{modint} satisfy
\begin{equation}
\label{lapTU}
\left[\varDelta_T+\tfrac12 s(1-s) \right] \mathcal{I} (s,\kappa)=\left[\varDelta_U+\tfrac12 s(1-s) \right] \mathcal{I} (s,\kappa ) = 12\, \sigma_1(\kappa)\, \delta_{s,1}\,,
\end{equation}
where the source term is due to the subtraction of the  simple pole in the integral \eqref{modint} at $s=1$. 

Moreover, it suffices to consider the integral $\cI (s) \equiv \cI  (s,1)$ since eqs. \eqref{NPHecke}, \eqref{DelTUtau} together with the self-adjointness of the Hecke operator allows one to express the  integral \eqref{modint} for general positive integer $\kappa$ as
\begin{equation}
\cI (s,\kappa ) =  H^{(T)}_\kappa \, \cI (s) = H^{(U)}_\kappa \, \cI (s)\,.
\end{equation}

\subsubsection{The BPS state sum representation}\label{BPSPoinca}

A novel procedure for evaluating the integral \eqref{modint} has been discussed in \cite{Angelantonj:2012gw}. It amounts to unfolding the  ${\rm SL} (2;\mathbb{Z})$ fundamental domain $\mathcal{F}$ against the absolutely convergent Niebur-Poincar\'e series. One thus arrives at the Schwinger-like representation
\begin{equation}
\cI (s ) = \sum_{\rm BPS} \,  \int_0^{\mathcal \infty} \frac{\de\tau_2}{\tau_2} \, 
 {\mathcal M}_{s,0} (- \tau_2 )\,  e^{-\frac{\pi\tau_2}{2} (|p_{\rm L}|^2 + |p_{\rm R}|^2)} \,,
\label{BPSint}
\end{equation} 
where $\sum_{\rm BPS}$ denotes a sum over integers $m_i, n^i$ subject to the `BPS constraint'\footnote{BPS states exist for any charges such that  $m_1 n^1 + m_2 n^2\leq 1$,
so \eqref{BPSconstr} picks only a subset of the allowed BPS states associated to the $q$-pole, which corresponds to the ground state of the non-supersymmetric side of the heterotic string.}
\begin{equation}
m_1 n^1 + m_2 n^2 = 1\ .
 \label{BPSconstr}
\end{equation}
 Performing the $\tau_2$-integral then yields \cite{Angelantonj:2012gw}
\begin{equation}
\cI (s) = \varGamma (s)\,  
\sum_{\rm BPS}\,\,  \left(\frac{ |p_{\rm L}|^2}{4} \right)^{-s}  \,
{}_2 F_1 \left(s, s; 2s \,;\, \frac{4 }{ |p_{\rm L}|^2} \right)\,.
\label{int2F1}
\end{equation}
The sum is absolutely convergent for ${\rm Re} (s)> 1$ and can be analytically continued to a meromorphic function on the complex $s$ plane with a simple pole at $s=1$ \cite{1004.11021, Angelantonj:2012gw}. The renormalisation prescription implies that this pole should be subtracted in defining the value of \eqref{int2F1} at $s=1$. Note that the hypergeometric function  reduces to simple elementary functions for integral positive values of $s$ \cite{Angelantonj:2012gw}.

Actually, eq. \eqref{int2F1} provides an explicit Poincar\'e series representation of the automorphic function $\cI_{2,2} (s,1)$ of ${\rm O} (2,2;\mathbb{Z})$. To illustrate this point, we note that the BPS constraint \eqref{BPSconstr} allows one to write
\begin{equation}
|p_{\rm L}|^2 = \frac{| \tilde {\bar T} - \tilde U |^2}{\tilde T_2\, \tilde U_2}\,,\qquad {\rm with}\quad 
\tilde T = \gamma_T \cdot T\,,
\qquad \tilde U = \gamma_U \cdot U \,,
\end{equation}
where $\gamma_T$ and $\gamma_U$ are both ${\rm SL} (2;\mathbb{Z})$ matrices. One may thus express the integral 
\begin{equation}
\cI (s) = \sum_{\gamma  \in \varGamma_{\rm diag} \backslash (\varGamma_T \times \varGamma_U) } \varphi \left( \frac{|\bar T - U|^2}{T_2 U_2}\right) \Bigg|\, \gamma \,. 
\end{equation}
as the ${\rm O} (2,2;\mathbb{Z})$ Poincar\'e series, with seed
\begin{equation}
\varphi (z) = \varGamma (s)\, \left( \frac{z}{4}\right)^{-s}\, {}_2 F_1 \left(s,s;2s; \frac{4}{z}\right) \,,
\label{seedBPS}
\end{equation}
invariant under the diagonal subgroup $\varGamma_{\rm diag}$ of $\varGamma_T \times \varGamma_U$.
It is straightforward to see that the action of the Hecke operator $H_\kappa \, \cI (s,1) = \cI (s,\kappa )$ amounts to rescaling  $|p_L|^2 \to |p_L|^2 /\kappa$ and  modifying the BPS constraint to $m_1 n^1 + m_2 n^2 = \kappa$.

The BPS-state-sum representation of the modular integral is thus manifestly invariant under the action of the duality group and is valid throughout the Narain moduli space. In particular, it makes manifest the  singularities of the integral \eqref{modint} in the vicinity of points of symmetry enhancement where $|p_{\rm R}|^2\to 0$ for some choice of the integers $m_i,n^i$ \cite{Angelantonj:2012gw}.  
If, however, one is interested in the asymptotic behaviour of \eqref{modint} at the boundary of the moduli space, one should instead resort to a Fourier-series representation of the integral. 
This is best accomplished at the level of the Schwinger-like integral \eqref{BPSint} as we now show.

\subsubsection{The Fourier series representation}

The BPS constraint \eqref{BPSconstr}
admits an infinite number of solutions which are in one-to-one correspondence with elements of ${\rm SL} (2;\mathbb{Z})$. For any given co-prime pair $(n^1 , n^2)$ the Euclidean algorithm provides a pair of co-prime integers $(m_1 ^\star, m_2^\star)$ solving \eqref{BPSconstr}. The most general solution can thus be obtained as
\be
\label{BPSsol}
m_1 =  m_1^\star + \tilde M n^2\, ,
\quad 
m_2 =  m_2^\star - \tilde M n^1\, ,
\ee
for any $\tilde M \in \mathbb{Z}$. Plugging this solution in \eqref{BPSint} and Poisson summing over $\tilde M$, provides the desired Fourier expansion of the integral,
\be
\label{intNPunfold}
\begin{split}
{\mathcal I} (s) =& \sum_{M\in\IZ}\, \sum_{(n^1,n^2)=1}\, \sqrt{T_2\,\tilde U_2}\  e^{2i \pi M (T_1 - \tilde U_1)}
\\
& \times \int_0^\infty \frac{\de\tau_2}{\tau_2^{3/2}} \, 
 {\mathcal M}_{s,0} (- \tau_2 )\,  \exp\left[-\pi\tau_2 \left( \frac{T_2}{\tilde U_2} + \frac{\tilde U_2}{ T_2}\right)
 -\frac{\pi M^2 T_2 \tilde U_2}{\tau_2} \right]
\end{split}
\ee
where
\be
\tilde U \equiv \tilde U_1 + i \tilde U_2 = \frac{m_1^\star U-m_2^\star}{n^2 U+n^1} \equiv \gamma\cdot  U\,,
\quad {\rm with}\qquad  \gamma=\begin{pmatrix} m_1^\star & -m_2^\star \\ n^2 & n^1 \end{pmatrix} \in {\rm SL} (2;\mathbb{Z})\,.
\ee
The integral in \eqref{intNPunfold} can be readily evaluated for any $T_2$ and $\tilde U_2$, using the explicit expression
\be
{\mathcal M}_{s,0} (t) = 2^{2s-1}\, \varGamma (s+\tfrac{1}{2})\, (4\pi |t|)^{\frac{1}{2}}\, I_{s-\frac{1}{2}} (2 \pi |t|) \,,\\
\ee
and eq. (16) of Section 8.6 in \cite{erdelyi}\footnote{We reproduce this formula for convenience:
$$
\int_0^{\mathcal \infty} \frac{\de t}{t} e^{-\frac{1}{2t}-\beta t} I_\nu(\gamma t)
= 2\, 
 K_\nu\left( \sqrt{\beta+\sqrt{\beta^2-\gamma^2}}\right) \,
 I_\nu\left( \sqrt{\beta-\sqrt{\beta^2-\gamma^2}}\right) \qquad (\beta >\gamma >0)\,.
 $$
 }, and reads
\be
{\mathcal I} (s) = {\mathcal I}^{(0)} (s) + {\mathcal I}^{(+)} (s) +  {\mathcal I}^{(-)} (s)\,,
\ee
with the zero-frequency mode given by
\begin{equation}
\cI^{(0)} (s) = 2^{4s-2}\, \sqrt{4\pi} \, \varGamma (s-\tfrac{1}{2}) \, \sum_{\gamma \in \varGamma_\infty \backslash \varGamma} (T_2 \tilde U_2)^s\, \left( T_2 +\tilde U_2 + |T_2 - \tilde U_2 |\right)^{1-2s} \,,
\end{equation}
the positive-frequency modes given by
\begin{equation}
\begin{split}
\cI^{(+)} (s) &= \sum_{M>0}\sum_{\gamma \in \varGamma_\infty \backslash \varGamma} \frac{e^{2i \pi M (T_1 - \tilde U_1 )}}{M} \, 
\\
&\qquad \times \cM_{s,0} \left( \tfrac{1}{2} \, M \left[T_2 + \tilde U_2 - |T_2 - \tilde U_2 | \right] \right)\, \cW_{s,0} \left( \tfrac{1}{2}\, M  \left[T_2 + \tilde U_2 + |T_2 - \tilde U_2 | \right]\right)\,,
\end{split}
\end{equation}
and the negative-frequency modes simply obtained via complex conjugation, $\cI^{(-)} (s) = [ \cI^{(+)} (s^\ast)]^\ast$. Notice that we have recognised the sum over the co-prime integers $(n^1 , n^2)$ as the sum over cosets  $\varGamma_\infty \backslash \varGamma$ where $\varGamma=\varGamma_U$.

Actually, the Fourier series expansion has different coefficients in the chambers $T_2>U_2$ and $T_2<U_2$. Without loss of generality we can always assume that both $T$ and $U$ lie in the fundamental domain $\cF$. Then, whenever $U$ satisfies $U_2 < T_2 $, so do all its images $\tilde U = \gamma U$, for any $\gamma \in \varGamma_\infty \backslash \varGamma$.
On the contrary, when $T_2>1$ all points $U$ with $U_2 > T_2$ are mapped to points $\tilde U$ with $\tilde U_2 < T_2$, except when $\gamma$ is in the coset of the identity.
If $T_2 <1$, then both $U$ and $\tilde U = -1/U$ may satisfy the inequality $\tilde U_2 > T_2$. 

As a result, the Fourier coefficients for $\cI (s)$ read
\begin{equation}
\cI^{(0)} (s) = 2^{2s}\, \sqrt{4\pi} \, \varGamma (s-\tfrac{1}{2}) \, \left[ T_2^{1-s}\, E (s,0;U) +\tfrac{1}{2} \sum_{\gamma = \mathbb{1}, S}  \varTheta (\tilde U_2 - T_2 )\,  \left( T_2^{s}\, \tilde U_2^{1-s} - T_2^{1-s}\, \tilde U_2 ^s \right) \right]\,,
\label{intzm}
\end{equation}
and 
\begin{equation}
\begin{split}
\cI^{(+)} (s) &= 2\, \sum_{M>0} \Biggl[ \frac{e^{2i\pi M T_1}}{M}\, \cW_{s,0} (M T_2 )\, \cF (s,M,0;U)
 + \sum_{\gamma = \mathbb{1}, S} \varTheta (\tilde U_2 - T_2 ) 
\\
&\quad \times  \frac{e^{2 i \pi M (T_1 - \tilde U_1 )}}{M}\, \left( \cM_{s,0} (MT_2 )\, \cW_{s,0} (M\tilde U_2 ) - \cW_{s,0} (M T_2) \, \cM_{s,0} (M \tilde U_2 )\right) \Biggr]\, \,.
\end{split}
\label{intpf}
\end{equation}
where $\varTheta (x)$ is the Heaviside function. Note that, upon  Fourier expanding the Niebur-Poincar\'e series, the expressions in the two chambers are related by $T\leftrightarrow U$ exchange. Indeed, one could have obtained the very same results by alternatively solving the BPS constraint in terms of pairs of co-prime momenta $(m_1 , m_2)$, which would result in a Fourier expansion with respect to $U$, convergent in the chamber $U_2>T_2$, with coefficients 
manifestly invariant under  ${\rm SL} (2;\mathbb{Z})_T$.

For convenience, we shall henceforth focus our attention to the fundamental chamber $T_2> U_2$, where the Heaviside function in \eqref{intzm} and \eqref{intpf} vanishes, and the above expressions simplify to
\begin{equation}
\begin{split}
\cI^{(0)} (s) &=  2^{2s}\, \sqrt{4\pi} \, \varGamma (s-\tfrac{1}{2}) \, T_2^{1-s}\, E (s,0;U)\,,
\\
\cI^{(+)} (s) &= 2\, \sum_{M>0} \frac{e^{2i\pi M T_1}}{M}\, \cW_{s,0} (M T_2 )\, \cF (s,M,0;U)\,.
\end{split}\label{intFourier}
\end{equation}
Remarkably, using \eqref{NPHecke}, the coefficient of the $M^{\rm th}$ Fourier mode can be seen to involve the Hecke transform of the same  function
$\cF(s,1,0)$ as the one appearing in the integrand of \eqref{modint}, except that the former
depends on $U$ while the latter depends on $\tau$. This is a manifestation of a fact first 
noticed in  \cite{Bachas:1997mc}, whose rationale will be explained in Section \ref{sec_unfoldlat}.

While the most relevant case  for heterotic string theory is $\kappa =1$, on which we have concentrated so far, the results above may be extended to higher values of $\kappa$ by acting with the Hecke operator. Specifically, one obtains
\begin{equation}
\begin{split}
\cI^{(0)} (s,\kappa ) &= 2^{2s} \, \sqrt{4\pi} \, \varGamma(s-\tfrac12) \, (\kappa T_2)^{1-s} \, \sigma_{2s-1}(\kappa)\,  E(s,0;U) \,,
\\
\cI^{(+)} (s,\kappa ) &= 2 \sum_{d|\kappa}  \sum_{M>0}\,  \frac{e^{2\pi\I d M T_1}}{M}\, \cW_{s,0}(d M T_2) \, \cF \left(s,\frac{\kappa M}{d},0;U \right) \,.
\end{split}
\end{equation}

\subsubsection{Unfolding against the Narain lattice \label{sec_unfoldlat}}

An alternative approach for obtaining the Fourier series representation of the modular integral, commonly used in the past string theory literature, is to unfold the fundamental domain against the Narain lattice \cite{O'Brien:1987pn,McClain:1986id, Dixon:1990pc}, rather than against the Niebur-Poincar\'e series. To this end, one resorts to the Lagrangian representation of the $(2,2)$ lattice and decomposes the sum of winding numbers  $A$ into orbits of the ${\rm SL} (2;\mathbb{Z})_{\tau}$ action. The vanishing orbit contribution, corresponding to the trivial matrix $A=0$, is absent in this case since the average of any Niebur-Poincar\'e series over the fundamental domain vanishes. The degenerate orbits, corresponding to non-vanishing matrices $A$ with $\det A =0$, precisely reproduce $\cI^{(0)}$ in \eqref{intFourier}.
The non-degenerate orbits, associated to the set of  matrices
\begin{equation}
A = \begin{pmatrix} k & j \\ 0 & p \end{pmatrix}\,, \qquad 0\le j < k \,, \quad p\not= 0\,,
\end{equation}
can all be reconstructed from the contribution of the unit matrix $A=\mathbb{1}$, via the action of the Hecke operator $H_{\det (A)}$ on the complex structure $U$. In fact,
one can write
\begin{equation}
\cI_{\rm non-deg} (s) = 4 \, T_2 \, {\rm Re}\, \sum_{M>0} e^{2i\pi M T} \, H_M^{(U)}\, \int_{\mathbb{C}_+} d\mu\, e^{- \frac{\pi M T_2}{\tau_2 U_2} |U-\tau |^2}\, \cF (s,1,0; \tau)\,,
\label{nondegint}
\end{equation}
and upon evaluating the integral over the upper-complex plane one reproduces the non-vanishing frequency contributions to the Fourier expansion \eqref{intFourier}. 
 
 In the framework of the old unfolding method,  the fact that the Fourier coefficients in \eqref{intFourier} involve the Hecke transform of the same function $\cF (s, 1 , 0)$ appearing in the integrand is a consequence of Fay's theorem \cite{zbMATH03547622}, which states that any eigenfunction $G (z)$ of the hyperbolic Laplacian with eigenvalue $s (1-s)$, not necessarily automorphic, has the mean value property 
\begin{equation}
\frac{1}{m (\rho ; s)} \int_{\mathbb{C}_+} d\mu \, \rho (r)\, G (z)  =  G (z_0 )\,.
\label{Fayth}
\end{equation}
Here $r$ is the geodesic distance between the points $z$ and $z_0$, and $\rho (r) $ is an arbitrary `measure' on the complex upper plane $\mathbb{C}_+$ that depends on $r$ but not on the polar angle between $z$ and $z_0$, and the weight factor $m (\rho ; s)$ is given by 
\begin{equation}
\label{defmrho}
m (\rho ; s) = 2 \pi \int_0^\infty \, P_s ( \cosh r) \, \rho (r) \, \sinh r \, dr\,,
\end{equation}
with $P_s (t)$ the Legendre function. The integral in \eqref{nondegint} is seen to be a special
case of \eqref{Fayth} with $z=\tau, z_0=U$ and
\begin{equation}
\rho (r) = e^{- \frac{\pi M T_2}{\tau_2 U_2} |\tau - U|^2} = e^{-4\pi M T_2 \sinh^2 \frac{r}{2}}  \,. \quad 
\end{equation}
The integral  \eqref{defmrho} can be computed with the help of 7.141.5 in \cite{GR}, 
\be
m (\rho ; s) = \frac{e^{- 2 \pi M T_2}}{M T_2}\, \cW_{s,0} (M T_2 )\ ,
\ee
which thus yields the result \eqref{intFourier}.

\subsection{The special cases of integral $s$}

Let us now focus on the cases  $s=1+n$, with $n$ a non-negative integer. For low enough values of $n$ the Niebur-Poincar\'e series $\cF(1+n,1,0)$ reduce to weak almost holomorphic modular forms, and their modular integral \eqref{modint} admits alternative representations. The case $s=1$ is special in that $\cF (1,1,0) = j (z )+24$ is holomorphic\footnote{Our convention for the Klein $j$-function does not involve a constant term in its Fourier expansion.}, and $\cI(1)$ is related to Borcherds' automorphic products. For $n$ large the Niebur-Poincar\'e series are genuinely non-holomorphic. However, as discussed in Section \ref{secNPS}, the linear combinations \eqref{lineardec} still reduce to weak almost holomorphic modular forms, and are thus amenable to the same type of treatment.

\subsubsection{The case $s=1$}

As already noted, for $s=1$ the integral \eqref{modint} has a pole which must be 
subtracted \cite{1004.11021, Angelantonj:2012gw}. 
This pole originates from  the non-holomorphic Eisenstein series $E(s,0;U)$ appearing in \eqref{intFourier}. Using the first Kronecker limit formula one thus obtains
\begin{equation}
\cI^{(0)} (1) = - 24 \, \log \left(T_2 U_2\, |\eta (U) |^4 \right) + {\rm const}\, ,
\label{s1zeromode}
\end{equation}
where the additive constant is computable with the renormalisation scheme of \cite{MR656029, Angelantonj:2011br}, but inconsequential for our purposes.

The positive frequency modes are regular at $s=1$ and, using $\cW_{1,0}(t)=e^{-2\pi| t|}$, they simplify to 
\begin{equation}
\begin{split}
\cI^{(+)} (1) &=2 \sum_{M>0} \frac{q_T^M}{M} \cF (1,M,0; U )
\\
&= 2 \sum_{M>0} \frac{q_T^M}{M} \left[ \cF (1,M,0; U ) - \tilde \cF_0 (1,M,0; U) \right] -48\, \log \, \left( q_T^{-1/24}\, \eta (T)\right) \,,
\end{split}
\end{equation}
where we have introduced the nome $q_T$ associated to the K\"ahler modulus $T$, and $\tilde \cF_0 (1,M,0) = 24\, \sigma_1 (M)$ is the constant zero-mode of Niebur-Poincar\'e series, see
\eqref{NPFouriermodes}.

Using holomorphy and modularity in the $T$ and $U$ variables, together with the behaviour $\cF (1,M,0 ) - \tilde \cF_0 (1,M,0) \sim q_U^{-M}$ at the cusp $\infty$, one may show
\begin{equation}
\sum_{M>0} \frac{q_T^M}{M} \left[ \cF (1,M,0; U ) - \tilde \cF_0 (1,M,0; U) \right] = - \log q_T \left( j (T) - j (U) \right)\,,
\end{equation}
which is simply a reformulation of the Borcherds' product formula \cite{zbMATH00220742}
\begin{equation}
 q_T^{-1} \, \prod_{M>0 , N\in \mathbb{Z}} \left( 1 - q_T^M \, q_U^{N} \right)^{c (NM)} = j (T) - j (U) \,,
\end{equation}
where $c (M)$ are the Fourier coefficients of the Klein $j$-function. 

Putting things together, the modular integral reads
\begin{equation}
\cI (1) = - 24 \, \log (T_2 U_2\, |\eta (T) \, \eta (U) |^4 ) - \log | j (T) - j(U) |^4 + {\rm const}\, .
\label{modint2}
\end{equation}
The first term on the r.h.s. originates from the constant in the decomposition $\cF (1,1,0) = j (z ) + 24$ and is the celebrated result of \cite{Dixon:1990pc}, whereas the second term originates from the Klein function and can be uniquely determined by harmonicity, automorphy and by the singularities of the modular integral \cite{LopesCardoso:1994ik, Harvey:1995fq}. These originate from additional massless states appearing at the special points $T = U$ and their ${\rm SL} (2;\IZ)$ images. Similar automorphic products arise for $\kappa>1$, and can be obtained by acting with the Hecke operators $H_{\kappa}$ in either $T$ or $U$ variables,  but have more complicated singular divisors.

\subsubsection{The case $s=1+n$ with $n>0$}

For integer $s=1+n >1$  the Niebur-Poincar\'e series $\cF (1+n, 1, 0)$ becomes weakly almost holomorphic. Using the fact 
\begin{equation}
\cW_{1+n,0} (t) = (-1)^n\, n!\, (4\pi |t|)^{-n}\, e^{- 2\pi | t|  }L_n^{(-1-2n)} (4 \pi |t| ) \,,
\end{equation}
with $L_n^{(\alpha )}$ being the associated Laguerre polynomials, we can express
the integral  in terms of the `combined polylogarithms'
introduced in \cite{Kiritsis:1997hf}
\begin{equation}
\textrm{L}_{(k)}(z)=\sum_{\ell=0}^{k}\frac{(k+ \ell )!}{\ell!(k-\ell )!(4\pi)^\ell } \, y^{k-\ell}\,\textrm{Li}_{k+\ell +1}(e^{2i \pi  z})\, .
\label{SVPL}
\end{equation}
To this end, notice that the non-holomorphic Eisenstein series may be expressed as
 \begin{equation}
E (1+n , 0; z) = y^{1+n} + \frac{ \sqrt{\pi} \, \varGamma (n+\frac{1}{2})\, \zeta (2n+1)}{n!\, \zeta (2n+2)}\, y^{-n} + 2 \, {\rm Re}\, \left[ \frac{\pi^{1+n}\, y^{-n}}{ n! \, \zeta (2n+2 )}\sum_{N>0}\, {\rm L}_{(n)} (Nz)\right]\,,
\label{EisenCombPolylog}
\end{equation}
so that $\cI^{(0)}$ can be easily written in terms of the single-valued polylogarithms. As for the non-vanishing modes, we recall the Fourier series expansion of the Niebur-Poincar\'e series at the special value $s=1+n$,
\begin{equation}
\cF (1+n,1,0 ) = \sum_{N\ge -1} F_n (N)\, (4\pi y)^{-n}\, L_n^{(-1-2n)} (4\pi N y)\, q^N\,,
\end{equation}
where
\begin{equation}
\begin{split}
F_n (-1) &= \varGamma (2n+2 ) \,, 
\\
F_n (0) &= \frac{(2 \pi)^{2+2n}\, (-1)^n}{\zeta (2n+2)}\,, 
\\
F_n (N>0) &= 4\pi (-1)^n\, \varGamma (2n+2 ) \, \frac{\cZ (N,-1 ; 1+n)}{N^n}\,,
\end{split}\label{NPFcoeffs}
\end{equation}
as can be derived from eq. \eqref{NPFouriermodes2} and Appendix A. 
Using \eqref{NPFouriermodes2}, one may check that the $F_n(N)$'s are in fact
the Fourier coefficients of 
$\cF(1+n,1,-2n)=\sum_{N\geq -1} F_n(N) q^N$.

Using the following expression for the product of two associated Laguerre polynomials
\begin{equation}
(n!)^2\, L_n^{(-1-2n)}(u )\, L_n^{(-1-2n)}(v)= 
\sum_{k=0}^n \sum_{\ell=0}^k \frac{(n+k)!\, (k+\ell)!}{k!\, (n-k)!\, \ell ! \, (k-\ell)!} (uv)^{n-k}\, (u+v)^{k-\ell}\,,
\end{equation}
one finds
\begin{equation}
\cI^{(+)} = \frac{2(-1)^n}{n!}\sum_{M>0}\sum_{N\in\mathbb{Z}}\, F_n (MN)\, \sum_{k=0}^n \frac{(n+k)!}{k!\,(n-k)!}\frac{(MN)^n}{(4\pi MN T_2 U_2 )^k}\, {\rm L}_{(k)} (MT + NU) \,,
\end{equation}
in agreement with \cite{Kiritsis:1997hf}.
Notice that the combination \eqref{SVPL} of the polylogarithms has the property that its real part is a single-valued function in the complex plane \cite{zbMATH04144378}. This ensures that the modular integral be a well-defined function in the fundamental Weyl chamber.

\section{Generalised  prepotentials for ${\rm SL}(2;\IZ)$  \label{sec_prepot}}

Using the behaviour \eqref{NPderiv} of Niebur-Poincar\'e series under the action of modular derivatives,   
we show in this section that for integer $s=1+n>1$, the modular integral \eqref{modint} can be expressed in terms of a holomorphic function  $f_n(T,U)$, which we refer to as the 
generalised prepotential\footnote{This equation holds for $n< 5$ and $n=6$, such that 
 $\cF(1+n,1,-2n)$ is a weak holomorphic modular form. For $n=5$ and $n\geq 7$, $\cI(1+n)$ can always 
 be integrated to a harmonic prepotential. For suitable linear combinations such that the 
 shadows of $\cF(1+n,\kappa,-2n)$ cancel, there is no obstruction in taking the prepotential 
 to be holomorphic.},
\begin{equation}
\cI (1+n) = 4\, {\rm Re}\, \frac{(- D_T D_U )^n}{n! }\, f_n (T,U) \ .
\label{ppotdef}
\end{equation}
The case $n=1$ arises in computations of one-loop  threshold corrections to gauge
couplings in heterotic string theory compactified on ${\rm K3}\times T^2$, where $f_1(T,U)$  is identified as the one-loop correction to the usual holomorphic prepotential $F(X)$  encoding the vector-multiplet self-couplings in  $\cN=2$ supergravity theories in four dimensions \cite{Antoniadis:1995ct,deWit:1995zg,Forger:1997tu}. The case $n=2$ arose in studies of $F^4$ and $R^4$ couplings in heterotic strings on $T^2$ \cite{Lerche:1998nx}, although the physical significance of the prepotential
$f_2$ in the context of eight-dimensional supergravity remains to be elucidated. The generalised
prepotentials $f_n$ were first introduced in \cite{Kiritsis:1997hf}, where integrals of the type \eqref{intintro} were considered with $\varPhi$ being an arbitrary weakly holomorphic modular form.  Since $\varPhi$ is in general a mixture of Niebur-Poincar\'e series $\cF(1+n,1,0)$ with different values of $n$, the integral \eqref{intintro} becomes a sum of iterated derivatives of various prepotentials $f_n$. One advantage of decomposing $\varPhi$ into a sum of Niebur-Poincar\'e series $\cF(1+n,1,0)$ is that it allows one to disentangle these contributions.

While it is interesting and useful that the integral  \eqref{modint} can be integrated to a holomorphic function, 
the drawback is that this function is ambiguous up to polynomials of degree $2n$ in $T$ and $U$ with real coefficients, as these polynomials lie in the kernel of ${\rm Re}(D_T D_U)^n$. This implies that under T-duality, the prepotential $f_n$ need not be a standard modular form of weight $-2n$, but rather can transform inhomogenously,  picking up extra polynomials under the action of 
the modular group. This behaviour is characteristic of Eichler integrals. Indeed, we shall see that
$f_n$ contains a $T$-independent term proportional to the Eichler integral of the holomorphic Eisenstein series $E_{2n+2}(U)$, which is responsible for this  modular anomaly. Alternatively, one  may express the integral as an iterated derivative of a {\it harmonic} function $h_n(T,U)$, which is modular covariant with  weight $-2n$ but not holomorphic. This is in fact our first step in establishing
 \eqref{ppotdef}.

\subsection{Generalised harmonic prepotentials}

To derive the harmonic prepotentials associated to the integral $\cI (1+n)$, we recall the action \eqref{NPderiv} and \eqref{EPderiv} of the modular derivatives on the Niebur and Eisenstein Poincar\'e series, and observe that
\be
\begin{split}
D^n \, q^{N} &= (-2N)^n\, \cW_{n+1,0}( N y)\, e^{2\pi\I N x} \,,
\\
D^n \, 1 &= \frac{2^{2n}}{\sqrt\pi}\,\varGamma(n+\tfrac12) \, ( - 2\pi y)^{-n} \,,
\end{split}
\label{obscura}
\ee
provided $q^M$ is treated as a generic mode in the Fourier-series expansion of a modular form of weight $-2n$ in $z$. Using \eqref{NPderiv} and \eqref{EPderiv} we may then cast our result \eqref{intFourier} for the integral as the action of the iterated derivatives
\begin{equation}
\cI (1+n) = 4\, {\rm Re}\, \frac{(- D_T D_U )^n}{n! }\, h_n (T,U )\,,
\end{equation}
with
\begin{equation}
h_n (T,U) =  (2\pi)^{2n+1}\,  E(1+n,-2n;U) + 
\sum_{N>0} \frac{2}{(2N)^{2n+1}}  \, q_T^{N}\, \cF(1+n,N ,-2n;U)\,.
\label{fnmaass}
\end{equation}
Since the Eisenstein and Niebur-Poincar\'e series appearing in \eqref{fnmaass} have $s=1-\frac{w}{2}$, the function $h_n(T,U)$ is harmonic in $U$, and  holomorphic in $T$, and we shall refer to it as the generalised harmonic  prepotential. For low values of $n$,
the non-vanishing frequency modes are in fact also holomorphic in $U$, but the zero-mode
part never is. Importantly,  $h_n$ is manifestly modular under the action of ${\rm SL} (2;\mathbb{Z} )_U$, with weight $-2n$. Yet, it is not modular under the full T-duality group,
in particular it is not invariant under $T\leftrightarrow U$. Note that $h_n$ is a priori
defined up to terms in the kernel of ${\rm Re}(D_T D_U)^n$. Our choice ensures that 
$h_n$ is modular under  ${\rm SL} (2;\mathbb{Z} )_U$.

As before, this result generalises to the case of $\kappa > 1$ by the action of the Hecke operator on either $T$ or $U$. Specifically,
\begin{equation}
h_n (T,U;\kappa ) = \kappa^{-n} \, H_\kappa \cdot h_n (T,U)\,.
\label{prepHecke}
\end{equation}
The overall factor is a consequence of the identity
\begin{equation}
\kappa^n \, H_\kappa \, D^n \, \cF (s,1,w) = D^n \, \cF (s,\kappa , w )\,.
\end{equation}

Notice, that generalised {\em harmonic} prepotentials can always be introduced independently of the value of $n$, even when the Niebur-Poincar\'e series in the integrand is no longer weak almost holomorphic but is rather genuinely non-holomorphic, which occurs when $n=5$ and $n>6$.

\subsection{Generalised holomorphic prepotentials and Eichler integrals\label{sec_eichler}}

As we have seen, the harmonic prepotential $h_n$ emerges rather naturally and has the advantage of being modular covariant, at the cost of sacrificing holomorphy.  Although this is legitimate from a mathematical viewpoint,  it is however difficult to reconcile with the holomorphic structure of $\mathcal{N}=2$ supergravity, since $h_1$ should compute the radiative correction to the holomorphic prepotential $F(X)$. As we shall see, this is not a problem and holomorphy may always be restored, whenever the Niebur-Poincar\'e series in the integrand {\em is} weakly almost holomorphic. 

For $n=1,2,3,4,6$ the only source of non-holomorphy in the harmonic prepotential is due to the zero-frequency mode $\cI^{(0)}$. As a result, we shall first restrict our analysis to this contribution. Using \eqref{EPderiv} one has
\begin{equation}
E (1+n,0) = \frac{(2\pi)^n}{n!}\, D^n \, E (1+n , -2n) = \frac{(2\pi)^n}{n!}\, {\rm Re}\, D^n \, E (1+n , -2n) \,,
\end{equation}
where $E (1+n ; -2n)$ is a harmonic Maa\ss\ form. As a result, it can be decomposed 
\begin{equation}
E (1+n ; -2n ) = \varPhi_{\rm h} + \varPhi_{\rm nh}\,,
\end{equation}
into its holomorphic (Mock modular) part $\varPhi_{\rm h} $ and the non-holomorphic complement $ \varPhi_{\rm nh}$, with
\begin{equation}
\varPhi_{\rm h}  (z) = \left( \frac{z}{2i}\right)^{2n+1} + \frac{(-1)^n\, \pi \, \zeta (2n+1)}{2^{2n+1}\, \zeta (2n+2)}
 +\frac{ (-1)^n\, \pi}{2^{2n}\, \zeta (2n+2)} \sum_{N>0} \sigma_{-2n-1} (N) \, q^N\,,
\end{equation}
and
\begin{equation}
\begin{split}
\varPhi_{\rm nh} (x,y) &= y^{2n+1} - \left( \frac{z}{2i}\right)^{2n+1} +  \frac{(-1)^n\, \pi \, \zeta (2n+1)}{2^{2n+1}\, \zeta (2n+2)} 
\\
&\quad +\frac{ (-1)^n\, \pi}{2^{2n}\, \zeta (2n+2)} \sum_{N>0} \sigma_{-2n-1} (N) \, \frac{\varGamma (2n+1 ; 4\pi N y )}{\varGamma (2n+1)} q^{-N}\,.
\end{split}
\end{equation}
The remarkable identities
\begin{equation}
\begin{split}
{\rm Re} \, \left[ D^n \left( \frac{z}{2i}\right)^{2n+1} \right] &= D^n \, y^{2n+1} \,,
\\
D^n \, \left[ \varGamma (2n+1;4 \pi N y)\, q^{-N} \right] &= \frac{(2n)!}{(4\pi N)^n}\, \left[ D^n \, q^N \right]^* \,,
\end{split}
\end{equation}
together with equations \eqref{obscura} yield the non-trivial result
\begin{equation}
D^n \, \varPhi_{\rm h} (z) = \left[ D^n \, \varPhi_{\rm nh} (x,y) \right]^* \, ,
\end{equation}
which originates from the fact that both the shadow and the ghost \cite{Angelantonj:2012gw} of the 
harmonic Maa\ss\ form $E(n+1,-2n)$ are proportional to the holomorphic Eisenstein series $E_{2n+2} (z)$. In fact, one can prove that
\begin{equation}
D^{2n+1}\, \varPhi_{\rm h} = \left[  (D^*)^{n+1} \, D^n \, \varPhi_{\rm nh} \right]^* = y^{-2-2n}\, \left[ \bar D \, \varPhi_{\rm nh}\right]^*\,,
\end{equation}
where the left-hand side defines the ghost and the right-hand side defines the shadow.

The above considerations allow one to write
\begin{equation}
\cI^{(0)} (n+1) = \frac{4}{n!} \, {\rm Re} \, (- D_T D_U )^n \, f_n^{(0)} \,,
\end{equation}
where 
\begin{equation}
\label{fn0eich}
f_n^{(0)} (U) = \alpha\, \tilde E_{-2n} (U)\,,
\end{equation}
is the zero-frequency mode of the desired generalised {\rm holomorphic} prepotential, written in terms of
\begin{equation}
\tilde E_{-2n} (z) = \frac{\zeta (2n+2 )}{2\pi i }\, z^{2n+1} + \frac{\zeta (2n+1)}{2} + \sum_{N>0} \sigma_{-1-2n} (N) q^N\,,
\end{equation}
and
\begin{equation}
\alpha = \frac{(-1)^n \, (2\pi )^{2n+2} }{2^{2n}\, \zeta (2n+2)}\,.
\label{alphaConst}
\end{equation}
One may recognise $\tilde E_{-2n}(z)$ as the  Eichler integral of the holomorphic Eisenstein series $E_{2n+2} (z)$ \cite{0990.11041}, defined by 
\be
E_{2n+2} (z) = \frac{2\pi\I}{(2n+1)! \, \zeta(2n+2)} \pa_z^{2n+1} \tilde E_{-2n}(z)\ .
\ee
As anticipated, this holomorphic contribution does not transform covariantly under the action of the modular group, but rather involves an inhomogeneous term, as we shall discuss in the following Section. Altogether, the generalised {\em holomorphic} prepotential reads
\begin{equation}
f_n (T,U) = \alpha\, \tilde E_{-2n} (U) + \sum_{N>0} \frac{2}{(2N)^{2n+1}}  q_T^{N}\, \cF(1+n,N ,-2n;U)\,.
\label{holpp}
\end{equation}
It is important to stress that this choice of holomorphic  prepotential is not unique and one has the freedom of adding a polynomial of degree $2n$ in $T$ and $U$ with real coefficients. In \eqref{holpp} we have made a convenient choice for this polynomial. 

Upon Fourier-expanding $\cF(1+n,1 ,-2n)=\sum_{N\geq -1} F_n(N) q^N$, the generalised holomorphic prepotential 
admits the alternative representation
\begin{equation}
f_n (T,U) = \alpha\, \tilde E_{-2n} (U) +2^{-2n}\, \sum_{N>0 \atop M\in \mathbb{Z}} F_n (NM)\, {\rm Li}_{2n+1} (q_T^N \, q_U^M )
\label{holppLi}
\end{equation}
in terms of polylogarithms. By further Fourier expanding the Eichler integral and rearranging terms, one arrives at
\begin{equation}
	\begin{split}
	f_n(T,U) =&\alpha\,\zeta(2n+2)\left[\frac{U^{2n+1}}{2\pi i}-\tfrac{1}{2}\mathcal{Z} (0,0; 1+n)\right]+\frac{F_n (-1)}{2^{2n}}\,{\rm Li}_{2n+1}\left(\frac{q_T}{q_U}\right)  \\
			& +\frac{1}{2^{2n}} \sum_{N,M\geq 0} F_n(NM)\,{\rm Li}_{2n+1}(q_T^N \,q_U^M) \,,
	\end{split}	
	\label{SL2ZprepPoly}
\end{equation}
which exposes the asymmetry of the generalised holomorphic prepotentials under $T\leftrightarrow U$ exchange, originating from the singularity at $T=U$ in the term proportional to  $F_n (-1)$.

Although the zero-frquency part $\cI^{(0)}$ may always be expressed in terms of a  holomorphic function $f^{(0)}_n$
for any $n$, this is no longer true for the non-vanishing frequency modes when $n=5$ and $n>6$, since the Niebur-Poincar\'e series $\cF (1+n,N , -2n)$ are genuine harmonic Maa\ss\ forms \cite{Angelantonj:2012gw}. This is related to the fact that, for these values of $n$, $\cF (1+n,1,0)$ is a genuine non-holomorphic function. However, one can always associate a generalised  holomorphic prepotential to the linear combination \eqref{lineardec}, representing weak almost holomorphic modular forms, since the shadows will cancel also in the prepotential.

Similarly to the harmonic case, the generalised holomorphic prepotential $f_n (T,U;\kappa )$ may be straightforwardly obtained via the action of the Hecke operator $H_\kappa$ as in eq. \eqref{prepHecke}.

\subsection{Modular properties of generalised holomorphic prepotentials\label{sec_modeich}}

As anticipated, the generalised holomorphic prepotential of eq. \eqref{holpp} does not transform homogeneously with weight $-2n$ under the action of the T-duality group. Rather, owing to the inherent polynomial ambiguity in $f_n$, one has
\be
\begin{split}
(c T+d)^{2n}\, (c' U+d')^{2n}\, 
f_n\left(\frac{a T+b}{c T+d}, 
\frac{a' U+b'}{c' U+d'} \right) =&  f_n(T,U) +  P_{\gamma,\gamma'}(T,U) \, ,
\\
f_n(T,U) =& f_n (U,T) + P_\sigma(T,U)\,,
\end{split}
\ee
where $P_{\gamma,\gamma'}(T,U)$ and $P_\sigma(T,U)$ are polynomials of 
degree $2n$ in $(T,U)$ with real coefficients. These polynomials themselves depend on the choice of
polynomial ambiguity in $f_n$, which we have fixed in the definition \eqref{holpp}. 
 It is important to note that there is no choice of this ambiguity that can trivialise all polynomials 
 $P$'s.

 As far as 
the action of ${\rm SL}(2;\IZ)_U$ is concerned, the anomaly can be traced to the presence of the 
Eichler integral $\tilde E_{-2n}(U)$ in the zero-th Fourier mode \eqref{fn0eich} with respect to $T$. Indeed,
 under a generic element $\gamma \in {\rm SL} (2; \mathbb{Z})$, 
 the Eichler integral  transforms as
\begin{equation}
(c z + d)^{2n}\, \tilde E_{-2n} \left( \frac{az+b}{cz+d}\right) = \tilde E_{-2n} (z) + P_\gamma (z)\,,
\label{eichler1}
\end{equation}
with $P_\gamma (z)$ being a degree $2n$ polynomial in the $z$ variable, that depends on the choice of $\gamma$. 
The polynomials $P_\gamma (z)$ are uniquely determined by $P_T (z)$ and $P_S (z)$, which are associated to the standard generators $T$ and $S$ of the modular group given in \eqref{defTS}. It is straightforward to derive $P_T$ since the nome $q$ is trivially invariant and the only source of inhomogeneity is the monomial $z^{2n+1}$. As a result,
\begin{equation}
P_T (z) = \frac{\zeta (2n+2 )}{2\pi i }\, \sum_{k=0}^{2n} {2n+1 \choose k} z^k \,.
\label{PT}
\end{equation}

The transformation under $S$ is more involved and  can be determined by the analytic properties of the $L$-series
\begin{equation}
L^\star (s) = \int_0^\infty d y\, y^{s-1} \sum_{N>0} \sigma_{-1-2n} (N) e^{-2\pi N y} = (2 \pi )^{-s} \, \varGamma (s)\, \zeta (s) \, \zeta (s+2n+1)\,,
\end{equation}
associated to $\tilde E_{-2n}$, as we derive in Appendix \ref{appEichler}. Notice that although the $L$-series satisfies the functional equation $s \to w-s$, with $w=-2n$, and has simple poles at $s=0$ and $s=-2n$ 
as in the case for  modular forms of weight $-2n$, the additional poles at $s=1-2p$ with $p=0, 1 , \ldots , n+1$ are actually responsible for the inhomogeneous transformation \eqref{eichler1}, and the associated residues uniquely fix
\begin{equation}
P_S (z) = - \frac{(2\pi i )^{2n+1}}{2} \sum_{k=1}^n \frac{B_{2k}\, B_{2 n - 2k +2}}{(2k)!\, (2n-2k+2)!}\,  z^{2k-1}\,,
\label{PS}
\end{equation}
thus generalising the result of \cite{0990.11041} to generic $n$.

Clearly, $P_T$ and $P_S$  determine the behaviour of the generalised holomorphic prepotential $f_n (T,U)$ under the action of ${\rm SL} (2; \mathbb{Z})_U$. Moreover, using the property of the polylogarithms
\begin{equation}
{\rm Li}_k (e^{2 i \pi z}) + (-1)^k \, {\rm Li}_k (e^{-2i\pi z} ) = - \frac{ (2\pi i )^k}{k!} B_k (z) \,, \qquad 0 < {\rm Re}\, (z) <1\,,
\label{analyticContPoly}
\end{equation}
with $B_k (z)$ the Bernoulli polynomials, one obtains
\begin{equation}
 P_\sigma (T,U) = - \frac{(2\pi i )^{2n+1}}{2^{2n}}\, \left[ B_{2n+1} (T-U) - T^{2n+1}+ U^{2n+1} \right] 
\,,
\label{Psigma}
\end{equation}
when $T$ and $U$ lie in the fundamental chamber, with real parts in the interval $(0,1)$. 
This expression generalises the result of \cite{Harvey:1995fq} to $n>1$.

The knowledge of the polynomials $P_S$, $P_T$ and $P_\sigma$ fully determines the transformation of the prepotential under the duality group, because any transformation  $g' \in {\rm SL} (2;\mathbb{Z})_T$ can be obtained by conjugating the associated transformation $g\in {\rm SL} (2;\mathbb{Z})_U$ by $\sigma$, {\em i.e.} $g' = \sigma^{-1} g \sigma$. However, it is important to note that, since $P_{\sigma^2} \not=0 $, the generator $\sigma$ is no longer an involution at the quantum level, and as a result, modular transformations $g$ and $g'$ no longer commute. The group generated by these transformations, which we call the monodromy group, is an extension of the classical T-duality
group ${\rm SL}(2;\IZ)_T\times {\rm SL}(2;\IZ)_U\times \sigma$ given by the braid group of $(X\times X)\backslash \varDelta$, where $X$ is the Riemann sphere with three punctures located at images of $e^{i\pi /2}$, $e^{i\pi/3}$ and $i\infty$ under the action of the Klein $j$ function, and $\varDelta$ is the diagonal divisor in $X\times X$. The braid group is generated by the elements $g_1 :\  U \to -1/U$, $g_2:\ U\to -1/(U+1)$ and $\sigma$, obeying the relations
\begin{equation}
g_1 ^2 = g_2^3 = \mathbb{1}\,, \qquad g_i \sigma g_j \sigma^{-1} = \sigma g_j \sigma^{-1} g_i\,.
\end{equation}
Using the results above, we find that the generators act on the  generalised holomorphic prepotential as
\begin{equation}
\begin{split}
f_n \left(T, -\frac{1}{U} \right) &= U^{-2n}\, \left[ f_n (T,U) + \alpha P_S (U) \right]\,,
\\
f_n \left(T,-\frac{1}{U+1}\right) &=(U+1)^{-2n} \, \left[ f_n (T,U) + \alpha \left(P_{T} (U) + P_S (U+1) \right) \right]\,,
\\
f_n (T,U) &= f_n (U,T ) + P_\sigma (T,U)\,,
\\
f_n \left(-\frac{1}{T},U\right) &= T^{-2n} \, \left[ f_n (T,U) + \alpha P_S (T) - P_\sigma (T,U) +T^{2n} \, P_\sigma \left( -\frac{1}{T}, U\right)  \right]\,,
\\
f_n \left(-\frac{1}{T+1},U\right) &= (T+1)^{-2n} \, \biggl[ f_n (T,U) + \alpha \left( P_S (T+1) + P_T (T) \right) 
\\
&\qquad\qquad\qquad\qquad - P_\sigma (T,U) +(T+1)^{2n} \, P_\sigma \left( -\frac{1}{T+1}, U\right)  \biggr]\,,
\end{split}
\end{equation}
with $P_T$, $P_S$ and $P_\sigma$ the polynomials given in eqs. \eqref{PT}, \eqref{PS} and \eqref{Psigma}.

For $n=1$, our analysis is in full agreement with the results of \cite{Harvey:1995fq, Antoniadis:1995ct}
 and extends it to higher values of $n$. The precise matching with \cite{Antoniadis:1995ct} at the level of the holomorphic prepotential is obtained by adding to $f_1 (T,U)$ the polynomial $3 + 5 U + 3 U^2$ that lies in the kernel of ${\rm Re} \, (-D_T D_U)$.

\subsection{Generalised Yukawa couplings \label{sec_yuk}}

In $\cN=2$ supergravity in four dimensions, the Yukawa couplings $Y_{ijk}$ are obtained from the third derivatives of the prepotential $F$. They are modular covariant, and, in the case of $i=j=k=T$ or $U$, holomorphic in $(T,U)$. Similarly, for heterotic strings compactified on $T^2$, the one-loop contributions to $F^4_{\mu\nu}$ couplings are fifth derivatives of the generalised prepotential $f_2$. In this subsection, we construct a similar set of functions $Y_{T^p U^q}$ with $p+q=2n+1$, which we
call generalised Yukawa couplings, which are proportional to $(2n+1)^{\rm th}$ derivatives of the generalised prepotential
$f_n$, independent of the polynomial ambiguity, holomorphic when $p=0$ or $q=0$, modular covariant for any $p,q$,  and related to one-loop modular integrals with momentum insertions. The physical significance of these functions is not clear at present, but we hope that they may provide hints towards a generalisation of the notion of special geometry to $n>1$.

We shall begin by studying the simplest instance of holomorphic generalised Yukawa couplings
\begin{equation}
Y_{T^{2n+1}} (T,U) = - 2^{-2n-1}\, D_T^{2n+1} \, f_n (T,U) =\left( \frac{\partial_T}{2\pi i}\right)^{2n+1}\, f_n (T,U) \,,
\end{equation}
where in the last equality we used Bol's identity  \cite{Bol}, which equates the $(1-w)$-th modular  derivative of a modular function of negative integer weight $w<0$ to the ordinary $(1-w)$-th derivative. Notice that, since the degree $2n$ polynomial in the inhomogeneous transformation of $f_n$ is annihilated by the derivatives, $Y_{T^{2n+1}}$ indeed transforms as a weight $(2n+2,-2n)$ holomorphic modular form with respect to ${\rm SL} (2;\mathbb{Z})_T \times {\rm SL} (2;\mathbb{Z})_U$, for $n$ sufficiently small. Inserting the expressions \eqref{holpp} and \eqref{holppLi}, one obtains
\begin{equation}
\begin{split}
Y_{T^{2n+1}} (T,U) &= 2^{-2n} \sum_{N>0} q_T^N\, \cF (1+n , N, -2n ; U)
\\
&= 2^{-2n} \sum_{N>0 \atop M\in \mathbb{Z}} N^{2n+1} \, F_n (MN)\, \frac{q_T^N \, q_U^M}{1-q_T^N \, q_U^M}\,,
\end{split}
\end{equation}
thus generalising the `multi-cover' formula for the standard Yukawa coupling $Y_{T^3}$ \cite{Candelas:1990rm}. This expression clearly exhibits a simple pole divergence at $q_T = q_U$,
\begin{equation}
Y_{T^{2n+1}} \sim \frac{(2n+1)!}{2^{2n}}\, \frac{q_T}{q_U - q_T}\,.
\end{equation}
This behaviour, together with holomorphy and modularity, uniquely fixes the expression of the Yukawa couplings,
\begin{equation}
\begin{split}
Y_{T^3} &= \frac{3}{2} \, \frac{E_4 (T) \, E_4 (U) \, E_6 (U)}{\varDelta (U)\, [ j(T) - j(U)]} \,,
\\
 Y_{T^7} &= \frac{315}{4} \, \frac{E_4^2 (T) \, E_6 (U)}{\varDelta (U)\, [ j(T) - j(U)]}\,,
\end{split}
\qquad
\begin{split}
Y_{T^5} &= \frac{15}{2} \, \frac{E_6 (T) \, E_4^2 (U) }{\varDelta (U)\, [ j(T) - j(U)]} \,,
\\
Y_{T^9} &= \frac{2835}{2} \, \frac{E_4 (T) \, E_6 (T) \, E_4 (U)}{\varDelta (U)\, [ j(T) - j(U)]} \,,
\end{split}
\end{equation}
which may be written compactly as
\begin{equation}
Y_{T^{2n+1}} (T,U) = 2^{-2n}\, \frac{E_{2n+2} (T) \, \cF (1+n,1,-2n;U)}{j (T) - j (U)}\,,
\label{YukawasSL2Z}
\end{equation}
in terms of the holomorphic Eisenstein and Niebur-Poincar\'e series with $s=1-\frac{w}{2}$.
Note that the $Y_{T^{2n+1}}$ have already been discussed in the literature in the context of heterotic string compactifications on ${\rm K3} \times T^2$ and on $T^2$ \cite{  Lerche:1998nx, 
Lerche:1999hg, Antoniadis:1995zn}, respectively. Similar results can be obtained for $Y_{U^{2n+1}}$ upon exchanging $T$ and $U$ in the above expressions. 

The holomorphic generalised Yukawa couplings can be related to modular integrals with lattice-momentum insertions. In fact, 
\begin{equation}
D_T^{n+1} \, \bar D _U^n \, \cI (1+n) = (-1)^{n+1} \frac{4 (2n)!}{n!}\, Y_{T^{2n+1}} (T,U) + \frac{2}{n!}\, D_T^{n+1} \, \bar D_U^n \,\left[ (- D_{ T} D_{ U} )^n f_n ( T ,  U )\right]^* \,,
\end{equation}
where we have used again Bol's identity  and the fact that $\bar D^n D^n $ is proportional to the identity operator when acting on Niebur-Poincar\'e series. The second term on the r.h.s. actually vanishes due to holomorphy of $f_n$ and of non-trivial combinatorial identities. Moreover, one can trade the covariant derivatives for insertions of lattice momenta to obtain
\begin{equation}
Y_{T^{2n+1}} = \frac{(-1)^{n+1}}{2^{4n+3}}\, \frac{U_2^n}{T_2^{n+1}}\, \int_\cF d\mu \, \tau_2 \, \langle p_{\rm L}^{2n+1}\, \bar p_{\rm R} \rangle \, \cF (1+n , 1 , -2n )\,.
\end{equation}
Here we have employed the notation 
\begin{equation}
\langle f(p) \rangle = \tau_2 \, \sum_{m_i,n^i} f (p) \, q^{\frac{1}{4} |p_{\rm L}|^2 }\, \bar q^{\frac{1}{4} |p_{\rm R}|^2 }
\end{equation}
for the Narain lattice with generic momentum insertion $f (p)$. 

Also in this case, the generalisation to higher values of $\kappa$ is straightforward and can be obtained via the action of the Hecke operator
\begin{equation}
Y_{T^{2n+1}} (T,U;\kappa ) = \kappa^{-n} \, H_\kappa \cdot Y_{T^{2n+1}} (T,U)\,.
\end{equation}

We can now move to discuss more general Yukawa couplings, which can be obtained as mixed derivatives of the generalised holomorphic prepotential. Aside from sporadic cases \cite{Lerche:1998nx, Foerger:1998kw, Lerche:1999hg}, these have not been extensively studied in the literature, and indeed there are ambiguities in their proper definition. In these cases Bol's identity is no longer applicable and modularity dictates a non holomorphic completion. For instance, in the combination
\begin{equation}
Y_{T^p U^q} (T,U;\kappa )= -  \frac{D_U^q}{ 2^{2n+1}}  \left[ D_T^p  f_n (T,U;\kappa )- \frac{(2\pi)^{2q-2}\, p!}{(2n)!\, (q-1)!} \frac{ U_2^{2n}}{T_2^{2n-2q+2}} \left( D_T^{q-1} D_U^{2n}\, f_n (T,U;\kappa )\right)^* \right] ,
\label{mixedY}
\end{equation}
with $p+q=2n+1$ and $1 \le q \le n <p$, the degree $2n$ polynomial in the inhomogeneous transformation of $f_n$ is exactly cancelled. After some tedious algebra, one may cast them as modular integrals with lattice momentum insertions
\begin{equation}
Y_{T^p U^q} = (-1)^{n+1} \frac{\kappa^q\, q!}{2^{4n -2p +1}}\, \frac{U_2^{n-q}}{T_2^{p-n}}\, \int_\cF d\mu \, \tau_2 \, \langle p_{\rm L}^{p-q}\, \bar p_{\rm R}\rangle \, \cF (1+n , \kappa , 2q-2n )\,.
\end{equation}
These mixed Yukawa couplings are clearly non-holomorphic functions, but have holomorphic weights $2p-2n$ and $2q-2n$ with respect to ${\rm SL} (2;\mathbb{Z})_T$ and ${\rm SL} (2;\mathbb{Z})_U$, respectively. 

Furthermore, observe that the quantity inside the brackets in eq. \eqref{mixedY} is itself modular covariant and may be expressed as the modular integral 
\begin{equation}
(-1)^n \frac{(2\pi)^{2q} \,\kappa^{q-n-1}\, (2p)!}{2^{4n-q+4}\, n!}\, \frac{U_2^n}{T_2^{p-n}} \int_\cF d\mu\, \tau_2^{q-1}\, \langle p_{\rm L}^p\, p_{\rm R}^{q-1}\rangle\, \cF (1+n, \kappa, q-p-1) \,.
\end{equation}
This is a non-trivial result, and allows one to straightforwardly compute the Fourier series expansions of such integrals by expressing them as derivative of the known generalised prepotential. 

\section{Modular integrals for $\varGamma_0 (N)$\label{HeckeIntegrals_sec}}

In perturbative computations in many string theory models,  one encounters one-loop integrals over the fundamental domain of various congruence subgroups rather than of the full modular group. This happens {\em e.g.} in orbifold compactifications involving non-factorisable tori \cite{Mayr:1993mq,Bailin:2014nna} or when supersymmetry is spontaneously broken, either partially or completely  \cite{Kiritsis:1998en, Kiritsis:1997ca, Angelantonj:2006ut, Angelantonj:2014dia, Faraggi:2014eoa}. In these cases, the one-loop amplitudes  can be expressed as
\begin{equation}
\mathcal{I} = \int_\cF d\mu \, \sum_{h,g} \mathcal{A} \left[{\textstyle{h\atop g}} \right] \,,
\label{OrbiInt}
\end{equation}
where the sum over $h$ runs over the untwisted and twisted orbifold sectors and the sum over $g$ imposes the orbifold projection. Clearly, the integrand is invariant under the full modular group, but each independent term in the sum  is typically invariant only under some congruence subgroup of ${\rm SL} (2;\mathbb{Z})$, except for the untwisted unprojected sector $(h,g)=(0,0)$ which is invariant under the full modular group. 
This suggests that an efficient way of evaluating \eqref{OrbiInt} is to  partially unfold the modular integral \cite{Angelantonj:2013eja}, {\it i.e.} collect the $(h,g)$ sectors into orbits of ${\rm SL}(2;\IZ)$ and keep one element in each orbit, now integrated over  the fundamental domain of the corresponding congruence subgroup. Schematically, 
\begin{equation}
\sum_{h,g} \mathcal{A} \left[{\textstyle{h\atop g}} \right] = \sum_{a} \sum_{\gamma \in \varGamma_a \backslash \varGamma } \mathcal{A}_a \, |\, \gamma
\end{equation}
where the amplitudes $\mathcal{A}_a $ are now invariant under the congruence subgroup $\varGamma_a \subset \varGamma$, and the sum over $a$ runs over the possible subgroups. As a result, 
\begin{equation}
\cI = \sum_a \cI_a = \sum_a \int_{\cF_a} d\mu \, \mathcal{A}_a\,,
\end{equation}
$\cF_a$ being the fundamental domain of $\varGamma_a$. The integrals $\cI_a$ can then be computed by employing similar methods as in the case of the full modular group \cite{Angelantonj:2013eja}. 

For simplicity, we shall restrict our analysis to the case of abelian $\mathbb{Z}_2$ and $\mathbb{Z}_3$ orbifolds, which play an important role in string vacuum constructions. 
These cases are special in that they involve a single Hecke congruence subgroup, $\varGamma_0 (2)$ and $\varGamma_0 (3)$, respectively. Other $\mathbb{Z}_M$ and non-abelian orbifolds can be dealt with similar methods, though they typically involve several orbits. 

We shall further focus on BPS-saturated amplitudes, where the orbifold
blocks are of the form
\be
\mathcal{A} \left[{\textstyle{h\atop g}}\right]=\frac{1}{N} \varGamma_{2,2} \left[{}^h_g \right] (T,U) 
\varPhi \left[{}^h_g\right] \,,
\ee
where $\varPhi \left[{}^h_g\right] $ is a weakly almost holomorphic modular form under $\varGamma_0(N)$ and 
\begin{equation}
\varGamma_{2,2} \left[ {}^h_g\right] (T,U,\tau) = \tau_2 \sum_{m_i , n^i \in \mathbb{Z}} e^{\frac{2\pi i g}{N} (\lambda_i n^i + \mu^i m_i )}\, q^{\frac{1}{4} |p_{\rm L} (h)|^2}\, \bar q^{\frac{1}{4} |p_{\rm R} (h)|^2} \,,
\end{equation}
is the shifted Narain lattice partition function with momenta
\begin{equation}
\begin{split}
p_{\rm L} (h) &= \frac{1}{\sqrt{T_2 U_2 }} \left[ m_2 (h) - U m_1 (h) + \bar T (n^1 (h) + U n^2 (h)) \right]\,,
\\
p_{\rm R} (h) &= \frac{1}{\sqrt{T_2 U_2 }} \left[ m_2 (h) - U m_1 (h) +  T (n^1 (h) + U n^2 (h)) \right]\,,
\end{split}
\label{shiftedmomenta}
\end{equation}
where $m_i (h) = m_i + \frac{h}{N} \lambda_i$, $n^i (h) = n^i + \frac{h}{N} \mu^i$. The integral shift vector $(\lambda_i , \mu^i)$  is constrained by modular invariance to satisfy $ \lambda_i \mu^i = 0 \ \mod N$.

For $\mathbb{Z}_N$ orbifolds with $N=2,3$,  partial unfolding leads to 
\begin{equation}
\cI = \frac{1}{N} \int_\cF d\mu \, \varGamma_{2,2}\left[ {}^0_0 \right] (T,U)\, \varPhi \left[{}^0_0\right] + \frac{1}{N} \sum_{g=1}^{N-1} \int_{\cF_N}d\mu \, \varGamma_{2,2} \left[{}^0_g \right] (T,U) \varPhi \left[{}^0_g\right] \,,
\end{equation}
where $\cF_N$ is the fundamental domain of $\varGamma_0 (N)$. The relation $\varPhi [{}^h_g] = \varPhi [{}^{N-h}_{N-g}]$, valid for conjugate orbifold sectors, allows one to write
\begin{equation}
\cI  = \frac{1}{N} \int_\cF d\mu \, \varGamma_{2,2}\left[ {}^0_0 \right]\, \left[ \varPhi \left[{}^0_0\right] - \sum_{\gamma \in \varGamma_0 (N)\backslash \varGamma} \varPhi [{}^0_1 ]\, \Big| \gamma \right] +
\int_{\cF_N}d\mu \, \varPhi \left[{}^0_1\right]\, 
\frac{1}{N} \sum_{g=0}^{N-1}  \varGamma_{2,2} \left[{}^0_g \right]  \,.
	\label{integralFullCongr}
\end{equation}
Although this way of writing may appear contrived, it is actually instrumental in the subsequent analysis, since the sum over $g$ enforces a $\mathbb{Z}_N$ projection on the momenta and/or windings.

The term in the square bracket is ${\rm SL} (2;\mathbb{Z})$ invariant, and therefore the first integral can be straightforwardly evaluated using the results of \cite{Angelantonj:2011br,Angelantonj:2012gw} and of Sections \ref{sec_intsl2z}, \ref{sec_prepot}. The second integral instead can be treated using the techniques in \cite{Angelantonj:2013eja}. Since any weak almost holomorphic modular form of $\varGamma_0 (N)$ can be uniquely decomposed in terms of a linear combination of Niebur-Poincar\'e series \cite{Angelantonj:2013eja}, we shall restrict our analysis to the prototype integral involving the shifted lattice times the Niebur-Poincar\'e series of $\varGamma_0 (N)$.

\subsection{Symmetries of the shifted Narain lattice}

To study the symmetries of the shifted Narain lattice, it is convenient to introduce the matrices
\begin{equation}
M = \begin{pmatrix} n^1 & m_2 \\ n^2 & - m_1
\end{pmatrix}
\,,
\qquad
\varLambda =  \begin{pmatrix} \mu_1 & \mu_2 \\ \lambda^2 & - \lambda^1
\end{pmatrix}\,,
\end{equation}
that encode the  momentum and winding numbers, and the shift vectors.  The shifted Narain lattice then reads
\begin{equation}
\begin{split}
\varGamma_{2,2} (T, U,\tau; \varLambda ) &\equiv \frac{1}{N}\sum_{g\in \mathbb{Z}_N} \varGamma_{2,2} \left[ {}^0_g\right] (T,U,\tau )
\\
&=\sum_{g \in \mathbb{Z}_N}\sum_{M\in {\rm Mat}_{2\times 2} (\mathbb{Z})}\frac{ e^{\frac{2 i \pi g}{N}\, {\rm tr}\, (M \, \varLambda )}}{N} \, 
e^{-2 i \pi \tau \, {\rm det} (M)} \, \exp \left[ -\frac{\pi \tau_2}{T_2 U_2} \left| \begin{pmatrix} 1 & U \end{pmatrix} M \begin{pmatrix}  T  \\ 1 \end{pmatrix} \right|^2 \right]\,.
\end{split}
\end{equation}

Its duality symmetry is clearly a subgroup of  ${\rm SL} (2;\mathbb{Z})_T \times {\rm SL}(2;\mathbb{Z} )_U \times \sigma$, which is the duality group for $\varLambda =0$, and can be easily determined as follows. A generic transformation in ${\rm SL} (2;\mathbb{Z})_T \times {\rm SL}(2;\mathbb{Z} )_U$ can be reabsorbed by the matrix redefinition
\begin{equation}
M \to \tilde M = \sigma_1 \gamma_U^t \sigma_1 M \gamma_T\,,
\end{equation}
where $\sigma_1$ is the Pauli matrix, and $\gamma_\alpha \in {\rm SL} (2;\mathbb{Z})_\alpha$, at the cost of changing the shift matrix
\begin{equation}
\varLambda \to \tilde \varLambda = \gamma_T^{-1} \, \varLambda \, \sigma_1 \gamma_U^{-t}  \,\sigma_1\,.
\end{equation} 
This transformation acts on the lattice as
\begin{equation}
\varGamma_{2,2} ( \gamma_T T , \gamma_U U ; \varLambda ) = \varGamma_{2,2} (T,U; \tilde\varLambda )
\end{equation}
and is a symmetry if and only if
\begin{equation}
\varLambda  =  \gamma_T^{-1} \, \varLambda \, \sigma_1 \gamma_U^{-t}\, \sigma_1 \qquad \mod\, N\,.
\label{dualityUT}
\end{equation}
Similarly, one can show that the involution $\sigma\, :\ T\leftrightarrow U$ is a symmetry if and only if
\begin{equation}
\varLambda = \sigma_1 \varLambda^t \sigma_1 \qquad \mod \, N\,.
\label{involTU}
\end{equation}
Actually, one has the option of combining $\sigma$ with any other involution which is in ${\rm SL} (2;\mathbb{Z})_T \times {\rm SL} (2;\mathbb{Z})_U$ but not in the perturbative duality group of $\varGamma_{2,2} (T,U;\varLambda )$.

As an example, let us consider the case of a momentum shift along the first direction, $\mu_i = 0 = \lambda^2$, $\lambda^1 =1$. Solving the eqs. \eqref{dualityUT} and \eqref{involTU} yields the duality group $\varGamma^1 (N) _T \times \varGamma_1 (N)_U \ltimes \sigma_{T\leftrightarrow -1/U}$, where the  involution is now given by $\sigma S$, where $S : z \to - 1/z$ is an order-two element which is in ${\rm SL } (2;\mathbb{Z})$ but not in $\varGamma_1 (N)$ or $\varGamma^1 (N)$. Notice, however, that any even-weight modular form of the congruence subgroup $\varGamma_1 (N)$ is automatically modular under the larger Hecke subgroup $\varGamma_0 (N)$. As a result, since the lattice has zero-weight with respect to the  $T$ and $U$ variables, the effective duality group is always enhanced and in this case reads $\varGamma^0 (N) _T \times \varGamma_0 (N)_U \ltimes \sigma_{T\leftrightarrow -1/U}$. Similarly, for the case of a winding shift along the second cycle of the $T^2$, with $\mu_2 =1$ and $\mu_1 = 0 = \lambda^i$, the duality group is $\varGamma_0 (N) _T \times \varGamma_0 (N)_U \ltimes \sigma_{T\leftrightarrow U}$.  Actually, the lattice partition function is now invariant under the permutations of $T$, $U$, $\tau$ 
\begin{equation}
\varGamma_{2,2}(T,U,\tau;\varLambda) = \varGamma_{2,2}(U,T,\tau;\varLambda) =\varGamma_{2,2}(T,\tau,U;\varLambda) 
\,,
\end{equation}
just like the unshifted lattice $\varGamma_{2,2} (T,U,\tau )$. 
An important by-product is the relation between the Atkin-Lehner \cite{0177.34901} involution $z'=-1/(Nz)$ in the $T$, $U$ and $\tau$ variables
\begin{align}
\varGamma_{2,2} \left(T,-\tfrac{1}{NU},\tau;\varLambda \right) = \varGamma_{2,2} \left(-\tfrac{1}{NT}, U ,\tau;\varLambda \right)
=\varGamma_{2,2} \left(T,U,-\tfrac{1}{N\tau};\varLambda\right) \,.
\end{align}

Before we conclude, we observe that different shift vectors can actually be mapped into each other by a suitable redefinition of the $T$ and $U$ moduli \cite{Gregori:1997hi}.


\subsection{Niebur-Poincar\'e series for $\varGamma_0 (N)$}

In this subsection we recall basic facts about modular forms and Poincar\'e series with respect to the  Hecke congruence subgroups $\varGamma_0 (N)$ with $N$ prime, referring the reader to the existing literature for more details \cite{iwaniec2, shimura, 0804.11039, Angelantonj:2013eja}. The fundamental domain $\cF_N = \varGamma_0 (N) \backslash \mathbb{H}$ possesses two cusps, at $z=0$ and $z=\infty$. These special points are characterised by their {\em width} $m_\mathfrak{a}$, which essentially counts the number of copies of the ${\rm SL} (2;\mathbb{Z})$ fundamental domain in $\cF_N$ that end at the cusp $z_\mathfrak{a}$. As a result, one finds $m_\infty =1$ while $m_0 = N$. The two cusps are mapped to each other via the Atkin-Lehner involution $z\to -1/(Nz)$.
For a given seed $f(z)$, invariant under $z\to z+1$, one defines the Poincar\'e series
attached to the cusp $\mathfrak{a}$ via 
\begin{equation}
F_\mathfrak{a} (z) = \sum_{\gamma \in \varGamma_\mathfrak{a} \backslash \varGamma_0 (N) } f( z) \, \Big|_w \, \sigma_\mathfrak{a}^{-1}\, \gamma\,,
\label{HeckPS}
\end{equation}
where $\big|_w$ denotes the Petersson slash operator,  $\varGamma_\mathfrak{a}$ is the stabiliser of the cusp $z_\mathfrak{a}$, and $\sigma_\mathfrak{a}$ are ${\rm SL} (2;\IR)$ matrices which relate $\varGamma_\mathfrak{a}$ to $\varGamma_\infty$, with
\begin{equation}
\sigma_\infty = \begin{pmatrix} 1 &0 \\ 0 & 1 \end{pmatrix} \,, \qquad
\sigma_0 = \begin{pmatrix} 0 & 1/\sqrt{N} \\ -\sqrt{N} & 0 \end{pmatrix} .
\end{equation}
A convenient choice of Poincar\'e series of $\varGamma_0 (N)$ consists in  
choosing the same seed $f(z)=\cM_{s,w} (- \kappa y)e^{-2 i \pi \kappa x}$ as for ${\rm SL} (2;\mathbb{Z})$ Niebur-Poincar\'e series. We shall denote them by $\cF_\mathfrak{a} (s,\kappa , w)$, and we shall refer to them as the $\varGamma_0 (N)$ Niebur-Poincar\'e series attached to the cusp $\mathfrak{a}$. They are absolutely convergent for ${\rm Re}\, (s) >1$, and can thus be used for unfolding the fundamental domain. The functions $\cF_\infty (s,\kappa , w) $ and $\cF_0 (s,\kappa , w)$ are actually related via the action of the Atkin-Lehner involution, namely
\begin{equation}
\cF_0 (s,\kappa , w; z) = N^{-w/2}\,  z^{-w}\, \cF_\infty (s,\kappa , w ; -1/Nz )\,.
\end{equation}
By construction, they are eigenmodes of  the weight-$w$ hyperbolic Laplacian with the same eigenvalue as in \eqref{laplEskw} and satisfy the same closure properties under modular differentiations as in \eqref{NPderiv}.  As a result, they are harmonic Maa\ss\ forms for $s= 1- \frac{w}{2}$, and turn out to be weakly holomorphic for low enough weight, when the shadow vanishes identically. The Hecke operators $H_n^{(N)}$ of  $\varGamma_0 (N)$ \footnote{
The Hecke operator $H_n^{(N)}$ of $\varGamma_0 (N)$ acts on the Fourier modes of a weight-$w$ modular form $\varPhi = \sum_m  \varPhi (m,y) \,e^{2i \pi m x}$ as
\begin{equation}
H^{(N)}_n \, \varPhi (m,y) = n^{1-w} \sum_{d|(m,n)} d^{w-1}\, \chi (d )\, \varPhi (nm/d^2 , d^2 y/n)\,, \nonumber
\end{equation}
with $\chi (d)$ the trivial Dirichlet character $\mod\ N$, such that $\chi (d) = 1$ if $(N,d)=1$ and $\chi (d) =0$ otherwise. 
It satisfies the commutative algebra $H^{(N)}_\kappa \, H^{(N)}_\lambda = \sum_{d|(\kappa , \lambda)} d^{1-w}\, \chi (d)\, H^{(N)}_{\kappa \lambda / d^2}$.
}
acts on the Niebur-Poincar\'e series as 
\begin{equation}
H^{(N)}_\kappa \, \cF_{\mathfrak{a}} (s,1,w;z) = \cF_{\mathfrak{a}} (s,\kappa , w;z) - \delta_{\mathfrak{a}\infty}\,\cF (s,\kappa /N,w;Nz )\,.
\end{equation}
The second term in the r.h.s. is understood to be present only when $\mathfrak{a}=\infty$ and $N|\kappa$. It is important to note that  $H^{(N)}_\kappa$  is self-adjoint with respect to the Petersson inner product  only when $(N,\kappa)=1$ \cite{shimura, 0804.11039}. 

The  Fourier expansion of the Niebur-Poincar\'e series $\cF_\mathfrak{a} (s,\kappa , w)$ around the two cusps can be found in \cite{Angelantonj:2013eja}. For small enough (negative) weight, in the holomorphic case $s=1-\frac{w}{2}$, one has 
\begin{equation}
\begin{split}
\cF_\mathfrak{a} \left(1-\tfrac{w}{2} , \kappa , w \right)\, |_w \, \sigma_\mathfrak{b} =& \delta_{\mathfrak{ab}} \, \frac{\varGamma (2-w)}{q^\kappa} + \frac{4\pi^2 \kappa}{(2\pi i \kappa)^w}\, Z_{\mathfrak{ab}} \left(0,\kappa ; 1-\tfrac{w}{2}\right)
\\
& + 4\pi \kappa  \varGamma (2-w) \sum_{m\ge 1 }^\infty \left( \frac{-m}{\kappa}\right)^{\frac{w}{2}}\, \cZ_{\mathfrak{ab}} \left( m , -\kappa ; 1 - \tfrac{w}{2}\right)  \, q^m\,,
\end{split}
\end{equation}
where $Z_{\mathfrak{ab}} (a,b;s)$ and $\cZ_{\mathfrak{ab}} (a,b;s)$ are the Kloosterman-Selberg zeta functions relating the two cusps $\mathfrak{a}$ and $\mathfrak{b}$ defined in appendix \ref{kloosterman}\footnote{Note that in the present work we employ a different notation for the Kloosterman-Selberg zeta function than the one of \cite{Angelantonj:2013eja}.}. 

As for the case of the modular group, the Niebur-Poincar\'e series $\cF_\mathfrak{a} (1-\frac{w}{2}+n,\kappa , w)$ can be used to decompose any generic $\varGamma_0 (N)$ modular form $\varPhi $ of negative weight as a linear combination
\begin{equation}
\varPhi (z) = \sum_{\mathfrak{a}=0,\infty} \sum_{-\kappa_\mathfrak{a} \le m < 0} \sum_{p=0}^n d_{\mathfrak{a} , p } (m)\, \cF_\mathfrak{a} \left(1 - \tfrac{w}{2}+p, m , w \right)\,.
\label{HCdecomp}
\end{equation}
The coefficients $d_{\mathfrak{a} , p } (m)$ are uniquely determined by matching the singular behaviour of both sides of the equation around the two cusps. Their explicit expression can be found in \cite{Angelantonj:2013eja}. 

We shall also need the  non-holomorphic weight-$w$ Eisenstein series $E_\mathfrak{a} (s,w)$  obtained by choosing the seed $f(z) = y^s$ in \eqref{HeckPS}. As in the ${\rm SL}(2;\IZ)$ case, they are eigenmodes of the hyperbolic Laplacian with the same eigenvalue as the Niebur-Poincar\'e series, absolutely convergent for ${\rm Re} (s) >1$, and  
 satisfy the same closure properties under modular differentiation as in \eqref{EPderiv}. For $s=1-\frac{w}{2}$ and $s=\frac{w}{2}$ the Eisenstein series become harmonic Maa\ss\ forms. For even weight $w>2$,  $E_{\mathfrak{a},w} \equiv E_\mathfrak{a} (\frac{w}{2},w )$ reduces to the conventional holomorphic Eisenstein series $E_w$ of $\varGamma_0 (N)$ attached to the cusp $\mathfrak{a}$. The Eisenstein series $E_\mathfrak{a} (s,w)$   can actually be expressed as linear combinations of the ${\rm SL} (2;\mathbb{Z})$ ones with rescaled arguments,
\begin{equation}
E_\infty(s,w;z)=\frac{N^{s+\frac{w}{2}}\,E(s,w;Nz)-E(s,w;z)}{N^{2s}-1}\,, \qquad
E_0(s,w;z)=\frac{N^s\,E(s,w;z)-N^{\frac{w}{2}}\,E(s,w;Nz)}{N^{2s}-1}\,.
\label{FuncRelatEisen}
\end{equation}
From these relations, the Fourier series expansions around the cusp $\mathfrak{b}$ of the $\varGamma_0 (N)$ Eisenstein series attached to the cusp $\mathfrak{a}$ can be straightforwardly extracted and one obtains 
\begin{equation}
\begin{split}
E_{\mathfrak a}(s,w;z) |_{\mathfrak{b}} &= \delta_{\mathfrak{ab}}\,y^{s-\frac{w}{2}}+i^{-w}\sqrt{\pi}\,\frac{\varGamma(s)\,\varGamma(s-\frac{1}{2})}{\varGamma(s+\frac{w}{2})\,\varGamma(s-\frac{w}{2})}\, Z_{\mathfrak{ab}}(0,0;s)\,y^{1-s-\frac{w}{2}} 
\\
&+ i^{-w}\pi^s (4\pi)^{\frac{w}{2}}\sum_{m\neq 0}\frac{1}{\varGamma(s+\frac{w}{2}{\rm sgn}(m))}\,|m|^{s+\frac{w}{2}-1}\, Z_{\mathfrak{ab}}(0,m;s)\,\mathcal W_{s,w}(my)\,e^{2\pi i m x} \,.
\end{split}
\end{equation}

\subsection{Properties of the $\varGamma_0 (N)$ modular integrals}

We can now proceed to the computation of the prototype integral
\begin{equation}
\cI_N (s,\kappa ; \mathfrak{a} ) = {\rm R.N.} \int_{\cF_N} d\mu \, \cF_\mathfrak{a} (s,\kappa , 0 )\, \varGamma_{2,2} (T,U ; \varLambda )
\label{HCprot}
\end{equation}
by unfolding $\cF_N$ against the Niebur-Poincar\'e series. Any $\varGamma_0 (N)$ integral involving the shifted Narain lattice times a generic weakly almost holomorphic modular form can be straightforwardly derived from \eqref{HCprot} by virtue of the decomposition \eqref{HCdecomp}.  The symbol R.N. stands for the $\varGamma_0 (N)$ invariant renormalisation prescription of \cite{Angelantonj:2013eja} for treating the infra-red divergences that may arise at the two cusps. Special care is needed for $s=1$ where the integral develops a simple pole, in which case, the renormalisation prescription amounts to its proper subtraction.

From the differential properties of the Niebur-Poincar\'e series and from the  identities
\begin{equation}
\varDelta_T \, \varGamma_{2,2} (T,U ; \varLambda) =
\varDelta_U \, \varGamma_{2,2} (T,U ; \varLambda) =
\varDelta_\tau \, \varGamma_{2,2} (T,U ; \varLambda) \,,
\end{equation}
obeyed by the shifted Narain lattice, for any choice of the shift $\varLambda$, it turns out that the $\varGamma_0 (N)$ modular integrals $\cI_N (s,\kappa ; \mathfrak{a})$ are themselves eigenmodes of both the (weight-zero) hyperbolic Laplacians $\varDelta_T$ and $\varDelta_U$ with eigenvalue $\frac{1}{2} s (s-1)$. For $s=1$ a constant source-term appears, similarly to the ${\rm SL} (2;\mathbb{Z})$ case \eqref{lapTU}. 

Whenever the shifted lattice $\varGamma_{2,2} (T,U ; \varLambda )$ is symmetric under the exchange of the modular parameter $\tau$ with either the $T$ and $U$ moduli, 
and provided that $(\kappa , N) =1$ so that the $\varGamma_0 (N)$ Hecke operator is self-adjoint,
one can restrict the analysis to the case $\kappa =1$ and $\mathfrak{a} = \infty$, since all other cases can be recovered via the action of the Hecke and Atkin-Lehner operators.

Indeed, following the steps outlined in Section \ref{sec_intsl2z} one finds
\begin{equation}
\cI_N (s , \kappa ; \mathfrak{a}) =  H^{(N)}_\kappa\cdot \cI_N (s ; \mathfrak{a})  \,,
\end{equation}
where the Hecke operator  acts on either $T$ or $U$, or on both, depending on the available symmetries and $\cI_N (s;\mathfrak{a}) \equiv \cI_N (s,1;\mathfrak{a})$.

Moreover, the integrals associated to the two cusps are not independent, but are related by the action of the Atkin-Lehner involution $\sigma_0$. In fact, since 
\begin{equation}
\cF_0 (s,\kappa , w) = \cF_\infty (s,\kappa , w) |\sigma_0\,,
\end{equation}
and since $\sigma_0$ is self-adjoint on the fundamental domain of $\varGamma_0 (N)$, one finds
\begin{equation}
\cI_N (s; \mathfrak{a}) =  \cI_N (s ; \infty ) |\sigma_\mathfrak{a}\,,
\end{equation}
where the Petersson slash operator can act on either variable, depending on the available symmetries. For $s=1+n$, this relation can be extended to the level of both the harmonic and holomorphic  prepotential 
\begin{equation}
f_n (T,U ;\mathfrak{a} ) = f_n ( T,U ; \infty ) |\sigma_\mathfrak{a} +\,  {\rm polynomial}\,,
\label{SlashPrepotCongr}
\end{equation}
due to the commutation property of the involution $\sigma_\mathfrak{a}$ with the modular derivative. Similarly, for $(\kappa,N)=1$, one may extend the relation \eqref{prepHecke} to both the harmonic and holomorphic prepotentials of $\varGamma_0(N)$ 
\begin{equation}
	f_n(T,U,\kappa;\mathfrak{a}) =\kappa^{-n}\,H^{(N)}_\kappa\cdot f_n(T,U;\mathfrak{a}) \,.
\end{equation}

\subsection{Evaluation of the modular integral $\cI_N (s; \mathfrak{a} )$}

For concreteness, we shall evaluate the integral \eqref{HCprot} for $\kappa =1$, and we shall consider explicitly only the case of a winding shift along the second cycle of the two-torus, with $\mu_2 =1$ and $\mu_1 = \lambda^i =0$, since the other cases can be treated similarly. In addition to the duality symmetry  $\varGamma_0 (N)_T \times \varGamma_0 (N)_U \times \sigma_{T\leftrightarrow U}$, inherited by the integral, the shifted lattice is also invariant under the triality symmetry $T\leftrightarrow U \leftrightarrow \tau$, which implies that the results for the cusp at zero, or for higher values of $\kappa$, can be obtained by acting with the Atkin-Lehner involution or Hecke operator on  $\cI (s;\infty)$. 

For the cusp at $\infty$ the shifted lattice $\varGamma_{2,2} (T,U; \varLambda )$ reduces to the conventional unshifted Narain lattice sum merely subjected to the constraint $n^2 = 0 \ \mod \, N$. One may thus readily obtain the result of the integral $\cI (s;\infty )$ from its ${\rm SL} (2;\mathbb{Z})$ cousin $\cI (s)$, both in the BPS-sum representation and in the Fourier-series one. 

Unfolding the fundamental domain against the Niebur-Poincar\'e series yields the BPS-state sum representation \cite{Angelantonj:2013eja}
\begin{equation}
\cI_{N} (s; \mathfrak{a}) = \frac{m_\mathfrak{a}}{N}\, \varGamma (s) \, \sum_{g=0}^{N-1} \sum_{{\rm BPS} (v_\mathfrak{a} g)} \, e^{2 i \pi u_\mathfrak{a} g n_2/N}\, {}_2 F_1 \left( s, s; 2s; \frac{4}{m_\mathfrak{a}\, |p_{\rm L} (v_\mathfrak{a} g)|^2}\right) \, \left( \frac{m_\mathfrak{a}\, |p_{\rm L} (v_\mathfrak{a} g)|^2}{4} \right)^{-s}\,,
\end{equation}
where the BPS-sum extends over integers $m_i, n^i$ subject to the `BPS-constraint'
\begin{equation}
m_\mathfrak{a}\, \left( |p_{\rm L} (v_\mathfrak{a} g)|^2 - |p_{\rm R} (v_\mathfrak{a} g )|^2\right) = 4\,,
\end{equation}
with the shifted left-moving and right-moving momenta given in eq. \eqref{shiftedmomenta}. The ratio $u_\mathfrak{a}/v_\mathfrak{a}$ identifies the coordinate of the cusp $\mathfrak{a}$ on the complex upper plane, {\em i.e.} $(u_0,v_0) = (0,1)$  and $(u_\infty,v_\infty) = (1,0)$. 

Following the same arguments as in Section \ref{BPSPoinca} , this expression can be cast as the Poincar\'e series 
\begin{equation}
\cI_{N} (s;\mathfrak{a} ) = \sum_{\gamma \in \varGamma_{\rm diag} \backslash (\varGamma_T \times \varGamma_U)} \left. \varphi \left( \frac{|\bar T - U |^2}{T_2 U_2}\right) \right| \, \sigma_{\mathfrak{a},i}^{-1} \, \gamma\,,
\end{equation}
where the seed $\varphi (z)$ is the same as in eq. \eqref{seedBPS}, and the involution $\sigma_{\mathfrak{a},i}$ acts on either of the $T$ and $U$ variables. For generic $N$ including $N=1$, $\varGamma_T \times \varGamma_U = \varGamma_0 (N)_T \times \varGamma_0 (N)_U$ and $\varGamma_{\rm diag} =  {\rm diag}  (\varGamma_0 (N)_T \times \varGamma_0 (N)_U)$.

This  representation is valid at any point in the Narain moduli space. It is tailored to exploit the analytic structure of the integral in the vicinity of points of gauge symmetry enhancement and provides a Poincar\'e series representation, thus making the duality symmetry manifest. As in the ${\rm SL} (2;\mathbb{Z})$ case, points of gauge symmetry enhancement arise at $p_{\rm R} (v_\mathfrak{a} g) =0$, {\rm i.e.} $T = \sigma_\mathfrak{a} \, U$ together with their $\varGamma_0 (N)$ images \cite{Angelantonj:2012gw}. Note that the singular locus depends on a choice of cusp.

The Fourier series representation depends on a choice of Weyl chamber. Although the structure of Weyl chambers depends on the value of $N$, in the chamber  $T_2 > U_2$ one finds
\begin{equation}
\cI_{N} (s ; \mathfrak{a} ) = \cI_{N}^{(0)} (s ; \mathfrak{a} ) +\cI_{N}^{(+)} (s ; \mathfrak{a} ) + \cI_{N}^{(-)} (s ; \mathfrak{a} ) \,, 
\end{equation}
with 
\begin{equation}
\begin{split}
\cI_{N}^{(0)} (s ; \mathfrak{a} ) &=2^{2s}\, \sqrt{4\pi} \, \varGamma (s-\tfrac{1}{2}) \, T_2^{1-s}\, E_\mathfrak{a} (s,0;U)\,,
\\
\cI_{N}^{(+)} (s ; \mathfrak{a} ) &= 2\, \sum_{M>0} \frac{e^{2i\pi M T_1}}{M}\, \cW_{s,0} (M T_2 )\, \cF_\mathfrak{a} (s,M,0;U)\,, 
\end{split}
\end{equation}
and $\cI^{(-)}_N (s;\mathfrak{a} ) = [ \cI^{(+)}_N (s ;\mathfrak{a})]^*$.

Notice that although these expressions are valid for $N$ prime, the choice $N=1$ reduces to eq. \eqref{modint}, if we assume that, in this case, $\mathfrak{a}$ labels the unique cusp $\infty$ of the modular group. In fact, in this case $\cF^{(1)}_\infty (s,1,0) \equiv \cF (s,1,0)$,  $\cF_1 \equiv  \cF$, while for the Narain lattice, any shift vector $\varLambda$ acts trivially by simply relabelling the integral momenta and windings, and thus $\varGamma_{2,2}^{(1)} (T,U;\varLambda ) \equiv \varGamma_{2,2} (T,U)$.

Of particular interest in string theory are the cases where $s=1+n$ is a positive integer. The Fourier series expansion then admits  alternative representations in terms of polylogarithms or, for $n=0$, can be summed in terms of combinations of the $\varGamma_0 (N)$ Hauptmodul, thus generalising the Borcherd's product formul\ae\ valid for ${\rm SL} (2;\mathbb{Z})$.

\subsubsection{The case $s=1$}

The case $s=1$ is special since the ${\rm R.N.}$ prescription requires the subtraction of a simple pole from the non-holomorphic Eisenstein series entering  the Fourier mode with vanishing frequency. Using the first Kronecker limit formula one obtains
\begin{equation}
\cI_N^{(0)}(1;\mathfrak{a})= -\tfrac{24}{(N-1)\nu_N}\,\log\left( T_2^{N-1} U_2^{N-1}\left|\left[ \frac{\eta^N(N U)}{\eta(U)} \Big| \sigma_{\mathfrak a} \right]\right|^4\right)+{\rm const}\,,
\end{equation}
where $\nu_N = {\rm vol} (\cF_N)/{\rm vol} (\cF)$ is the index of $\varGamma_0 (N)$ inside ${\rm SL} (2;\mathbb{Z})$.
Notice that this expression is valid also for $N=1$, if we allow the formal limit $N\to 1$. In fact, $\eta (N U) \sim \eta (U) + \cO (N-1)$, and thus
\begin{equation}
\lim_{N\to 1} -\frac{24}{(N-1)\nu_N} \log \, \left( T_2^{N-1} U_2^{N-1}\left| \frac{\eta^N(N U)}{\eta(U)} \right|^4\right) = -24 \, \log \left(T_2 U_2 \, |\eta (U) |^4\right)\,,
\end{equation}
in agreement with eq. \eqref{s1zeromode}. 
The positive-frequency modes are regular at $s=1$ and read
\begin{equation}
\begin{split}
\cI_N^{(+)}(1;\mathfrak{a}) &= 2\sum_{M>0}\frac{q_T^M}{M}\,\mathcal{F}_{\mathfrak{a}}(1,M,0) 
\\
&=2\sum_{M>0}\frac{q_T^M}{M}\,\left[\mathcal{F}_{\mathfrak{a}}(1,M,0)-\tilde{\mathcal{F}}_{\mathfrak{a}}^{(0)}(1,M,0)\right]
+4\pi i T\,\delta_{\mathfrak{a}\infty} 
-\tfrac{24}{(N-1)\nu_N}\,\log\left( \left[ \frac{\eta^N(NT)}{\eta(T)} \Big| \sigma_{\mathfrak{a}}\right]\right)^2 \,,
\end{split}
\end{equation}
where $\tilde{\mathcal{F}}_{\mathfrak{a}}^{(0)}(1,M,0)=4\pi^2 M\, Z_{\mathfrak{a}\infty}(0,M;1)$ is the  constant zero-mode of the Niebur-Poincar\'e series of $\varGamma_0(N)$ attached to the cusp $\mathfrak a$.

Using the properties of holomorphy and modularity in the $T$ and $U$ variables, together with the singularity structure of $\mathcal{F}_{\mathfrak{a}}(1,M,0)-\tilde{\mathcal{F}}_{\mathfrak{a}}^{(0)}(1,M,0)$ around the two cusps, one obtains the identity
\begin{equation}
\sum_{M>0}\frac{q_T^M}{M}\left[\mathcal{F}_{\mathfrak{a}}(1,M,0)-\tilde{\mathcal{F}}_{\mathfrak{a}}^{(0)}(1,M,0)\right]+2\pi iT\,\delta_{\mathfrak{a}\infty} = -\log\frac{\left( j_{\infty}(T)-j_{\mathfrak{a}}(U)\right)}{\left( j_\infty(T)-\frac{24}{N-1}\right)^{\delta_{\mathfrak{a}0}}} \,,
\label{Borch1congr}
\end{equation}
where $j_{\mathfrak{a}}(z)=\sum_{M}c_{\mathfrak{a}}(M)\,q^M$. In fact, $j_\infty(z)$ is the Hauptmodul  for $\varGamma_0(N)$, while 
$j_0(z)=j_\infty(-1/Nz)$ is its image under the Atkin-Lehner involution.
Eq. \eqref{Borch1congr} generalises the Borcherds' product formula \cite{zbMATH00220742} to the case of $\varGamma_0(N)$ congruence subgroups
\begin{equation}
\prod_{K>0,\,L\in\mathbb Z}\left(\frac{(1-q_T^K \,q_U^L)^N}{1-q_T^{NK}\,q_U^{NL}}\right)^{c_\mathfrak{a}(KL)}= \frac{\left( j_\infty(T)-j_{\mathfrak{a}}(U)\right)^N}{\left(j(NT)-j(NU)\right)^{\delta_{\mathfrak{a}\infty}}\left(j_\infty(T)-\frac{24}{N-1}\right)^{N\delta_{\mathfrak{a}0}}} \,.
\end{equation}

Putting things together, the modular integral reads
\begin{equation}
\begin{split}
\cI_N (1; \mathfrak{a} ) =& -\tfrac{24}{(N-1)\nu_N}\,\log\left( T_2^{N-1} U_2^{N-1}\left|
\left[\frac{\eta^N(NU)}{\eta(U)}\, \Big| \, \sigma_{\mathfrak a}\right]
\,
\left[\frac{\eta^N(NT)}{\eta(T)} \, \Big| \sigma_{\mathfrak a}\, 
\right]
\right|^4\right)  
\\
& -\log\frac{\left| j_{\infty}(T) - j_{\mathfrak{a}}(U)\right|^4}{\left|j_\infty(T)-\frac{24}{N-1}\right|^{4\delta_{\mathfrak{a}0}}} +{\rm const}\,.
\end{split}
\label{congrResult1}
\end{equation}
The first term on the r.h.s. originates from the constant term  in the decomposition $\cF_\mathfrak{a} (1,1,0) = j_\mathfrak{a} (z) + \tilde{\mathcal F}_{\mathfrak{a}}^{(0)}(1,1,0)$, whereas the second term originates from the $\varGamma_0 (N)$ Hauptmodul. By employing the identity
\begin{equation}
	\left( j_0(T)-j_\infty(U)\right)\left(j_\infty(T)-\tfrac{24}{N-1}\right) = \left( j_0(U)-j_\infty(T)\right)\left(j_\infty(U)-\tfrac{24}{N-1}\right) \,,
\end{equation}
it is straightforward to verify that the expression \eqref{congrResult1} is indeed invariant under the duality symmetry $\varGamma_0(N)_T\times\varGamma_0(N)_U \ltimes \sigma_{T\leftrightarrow U}$. The modular integral \eqref{congrResult1} develops logarithmic singularities at the point $T=U$ and its $\varGamma_0(N)$ images, ascribed to the presence of extra massless states that would arise at symmetry enhancement points in the Narain moduli space.

Before closing this Section, we shall also give the expressions for the harmonic integrals involving the $\varGamma_0(N)$ Hauptmodul for $N=2,3$. Using the result
\begin{equation}
	{\rm R.N.}\int_{\mathcal F_N} d\mu\,\varGamma_{2,2}[^0_1] = -\frac{1}{N-1}\,\log (T_2 U_2)^{N-1}\,\left| \frac{\eta^N(NT)\,\eta^N(NU)}{\eta(T)\,\eta(U)}\right|^4 +{\rm const}\,,
\end{equation}
for integrals of the shifted Narain lattice established in \cite{Angelantonj:2013eja}, together with eq. \eqref{integralFullCongr}, we obtain
\begin{equation}
	\begin{split}
	{\rm R.N.} \int_{\mathcal F_N} d\mu\,\varGamma_{2,2}[^0_1]\,j_\infty = & -\frac{N}{N-1}\,\log\left| \frac{j_\infty(T)-j_\infty(U)}{\left(j_\infty(T)-\frac{24}{N-1}\right)\left(j_\infty(U)-\frac{24}{N-1}\right)}\right|^4 \\
	&+ \frac{1}{N-1}\,\log\left|j(T)-j(U)\right|^4 +{\rm const} \,,
	\end{split}
\end{equation}
and
\begin{equation}
	{\rm R.N.} \int_{\mathcal F_N} d\mu\,\varGamma_{2,2}[^0_1]\,(j_0-\tfrac{24}{N-1}) =  -\frac{N}{N-1}\,\log\left| j_0(T)-j_\infty(U)\right|^4 +{\rm const} \,,
\end{equation}
which generalise the ${\rm SL}(2;\mathbb Z)$ result \eqref{modint2} first obtained in \cite{Harvey:1995fq} to the $\varGamma_0(N)$ integrals. These results have been used in \cite{Angelantonj:2014dia} to derive the universal behaviour of gauge thresholds in stable non-supersymmetric heterotic vacua.


\subsubsection{The case $s=1+n$ with $n>0$}

For integer $s =1+n>1$, and for small enough $n$, the Niebur-Poincar\'e series $\cF_\mathfrak{a}^{(N)} (1+n,1,0)$ becomes almost holomorphic and admits the Fourier series expansion at the cusp $\infty$
\begin{equation}
\cF_\mathfrak{a}^{(N)} (1+n,1,0) = \sum_m F_{\mathfrak{a},n} (m) \, (4\pi y)^{-n}\, L_n^{(-2n-1)} (4 \pi m y )\, q^m\,,
\end{equation}
in terms of the associated Laguerre polynomials, where $F_{\mathfrak{a},n}$ are the Fourier coefficients of $\cF_\mathfrak{a}(1+n,1,-2n)$ at the cusp at infinity,
\begin{equation}
\begin{split}
F_{\mathfrak{a},n}(-1) &= \varGamma(2n+2)\,\delta_{\mathfrak{a}\infty} \,,
\\
F_{\mathfrak{a},n}(0) &= (2\pi)^{2n+2}\, (-1)^n\, Z _{\mathfrak{a}\infty}(0,1;n+1) \,, 
\\
F_{\mathfrak{a},n}(L>0) &= 4\pi\,(-1)^n\,\varGamma(2n+2)\,\frac{\mathcal Z_{\mathfrak{a}\infty}(L,-1;n+1)}{L^n} \,.
\end{split}
\end{equation}

Similarly to the ${\rm SL}(2;\mathbb Z)$ case, the integral may be expressed in terms of the single-valued polylogarithms of eq. \eqref{SVPL}. One finds
\begin{equation}
\begin{split}
\cI_N^{(+)} (1+n;\mathfrak{a}) &= \frac{\delta_{\mathfrak{a\infty}}}{N} \, \cI^{(+)} (NT, NU) 
\\
&\quad + \frac{2\, (-1)^n}{n!} \sum_{M>0\atop L\in \mathbb{Z}} F_{\mathfrak{a},n} (LM) \, \sum_{k =0}^n \frac{ (n+k)!}{k!\,(n-k)!} \, \frac{(ML)^n}{(4 \pi ML T_2 U_2)^k}\, \hat {\rm L}^{(N)}_{(k)} (MT + LU) \,,
\end{split}
\end{equation}
where we have defined the  `combined $\varGamma_0(N)$  polylogarithms'
\begin{equation}
	\hat{\rm L}^{(N)}_{(k)}(z) \equiv {\rm L}_{(k)}(z) - \frac{1}{N^{2k+1}}\, {\rm L}_{(k)}(Nz) \,.
\end{equation}
The zero mode $\cI_N^{(0)}$ similarly admits a representation in terms of single-valued polylogarithms, by using \eqref{EisenCombPolylog} together with the functional relations \eqref{FuncRelatEisen} in order to express the non-holomorphic Eisenstein series $E_\mathfrak{a}(1+n,0;z)$ of $\varGamma_0(N)$ in terms of the ${\rm SL}(2;\mathbb Z)$ ones, $E(1+n,0;z)$ and $E(1+n,0;Nz)$. 


\subsection{Generalised holomorphic prepotentials for $\varGamma_0(N)$}

As in the case of the full modular group, for integer $s=1+n>1$ one may define generalised harmonic and holomorphic prepotentials also for the $\varGamma_0(N)$ integral $\cI_N$ via
\begin{equation}
\begin{split}
\cI_N(1+n;\mathfrak{a}) &= 4\,{\rm Re} \frac{(-D_T D_U)^n}{n!}\, h_{n}(T,U;\mathfrak{a}) \,,
\\
&= 4\,{\rm Re} \frac{(-D_T D_U)^n}{n!}\,f_{n}(T,U;\mathfrak{a}) 
\,.
\end{split}
\end{equation}
The expression of the integral in terms of iterated derivatives of the prepotential is again possible due to the closure property of Niebur-Poincar\'e series under modular differentiation. 

The prepotentials for $\varGamma_0(N)$ can be obtained by an analysis parallel to the ${\rm SL}(2;\mathbb Z)$ case and one obtains
\begin{equation}
	h_{n}(T,U;\mathfrak{a}) = (2\pi)^{2n+1}\,E_\mathfrak{a}(n+1,-2n;U) + \sum_{M>0}\frac{2}{(2M)^{2n+1}}\,q_T^M\,\mathcal{F}_{\mathfrak{a}}(1+n,M,-2n;U) \,,
\end{equation}
for the generalised harmonic and
\begin{equation}
	f_{n}(T,U;\mathfrak{a}) = \frac{(-1)^n\,(2\pi)^{2n+2}}{2^{2n}\,\zeta(2n+2)}\,\tilde{E}_{-2n,\mathfrak{a}}(U) +\sum_{M>0}\frac{2}{(2M)^{2n+1}}\,q_T^M\,\mathcal{F}_{\mathfrak{a}}(1+n,M,-2n;U) \,,
	\label{HolPrepotCongr1}
\end{equation}
for the generalised holomorphic prepotentials. Here, $\tilde E_{-2n,\mathfrak{a}}(U)$ is the $\varGamma_0(N)$ Eichler integral  of weight $-2n$ associated to the holomorphic Eisenstein series $E_{2n+2,\mathfrak{a}} $, attached to the cusp $\mathfrak{a}$. Its Fourier expansion about a cusp $\mathfrak{b}$ reads
\begin{equation}
\tilde{E}_{-2n,\mathfrak{a}}(z) \big| \sigma_{\mathfrak{b}} = \zeta(2n+2) \left[ \frac{z^{2n+1}}{2\pi i}\,\delta_{\mathfrak{ab}}+\tfrac{1}{2} Z_{\mathfrak{ab}}(0,0;n+1)+\sum_{M>0} Z_{\mathfrak{ab}}(0,M;n+1)\,q^M \right]\,.
\end{equation}
The $\varGamma_0(N)$ Eichler integral $\tilde{E}_{-2n,\mathfrak{a}}$ inherits the functional relation properties of the Eisenstein series $E_{\mathfrak{a}}(n+1,-2n;z)$ and may be expressed in terms of ${\rm SL}(2;\mathbb Z)$ Eichler integrals via
\begin{equation} 
	\tilde{E}_{-2n,\infty}(z) = \frac{ N\,\tilde{E}_{-2n}(Nz)-\tilde{E}_{-2n}(z)}{N^{2n+2}-1} \,,\qquad \tilde{E}_{-2n,0}(z) = \frac{ N^{2n+1}\,\tilde{E}_{-2n}(z)-\tilde{E}_{-2n}(Nz)}{N^n(N^{2n+2}-1)} \,.
	\label{EichlerCongrFuncRel}
\end{equation}

Of course, since holomorphic prepotentials are only defined up to terms in the kernel of ${\rm Re}(-D_T D_U)^n$, the expression \eqref{HolPrepotCongr1}  for $f_n(T,U;\mathfrak{a})$ corresponds to a specific choice for the polynomial ambiguity. Notice that, due to the inhomogeneous transformation of the Eichler integral, the action of $\sigma_{0}$  now relates the holomorphic prepotential $f_n(T,U;0)$ attached to the cusp $\mathfrak{a}=0$ to the prepotential $f_n(T,U;\infty)$ attached to the cusp at $\infty$ with an additional polynomial dependence in the $U$ variable, so that eq. \eqref{SlashPrepotCongr} becomes
\begin{equation}
	f_n(T,U;0) = f_n(T,U;\infty)|\sigma_0-\alpha\,\frac{N^{2n+1}\,P_S(U)-P_S(NU)}{N^n\,(N^{2n+2}-1)} \,,
\end{equation}
where $P_S(z)$ is the polynomial transformation of the ${\rm SL}(2;\mathbb Z)$ Eichler integral given in \eqref{PS}  and $\alpha$ is the constant defined in \eqref{alphaConst}.

In terms of the Fourier coefficients $F_{\mathfrak{a},n}(M)$ of the Niebur-Poincar\'e series $\mathcal F_{\mathfrak{a}}(1+n,1,-2n)$, we may cast the generalised holomorphic prepotential in the alternative representation
\begin{equation}
\begin{split}
f_n(T,U;\mathfrak{a}) = & \frac{\delta_{\mathfrak{a}\infty}}{N^{2n+1}}\,f_n(NT,NU) + \frac{F_{\mathfrak{a},n}(-1)}{2^{2n}} \hat{{\rm Li}}_{2n+1}^{(N)}\left(\frac{q_T}{q_U}\right) 
\\
&+ \frac{1}{2^{2n}}\sum_{M,L\geq 0} F_{\mathfrak{a},n}(ML)\,\hat{\rm Li}_{2n+1}^{(N)}(q_T^M\,q_U^L)  
\\
&+ \tfrac{1}{2}\alpha\, \zeta(2n+2)\Bigr[ Z_{\mathfrak{a}\infty}(0,0;n+1)-\delta_{\mathfrak{a}\infty} Z(0,0;n+1) 
\\
& -2(1-N^{-2n-1})\zeta(2n+1) Z_{\mathfrak{a}\infty}(0,1;n+1)\Bigr] \,,
\end{split}	
\label{HolPrepotCongr2}
\end{equation}
where $f_n(T,U)$ is the ${\rm SL}(2;\mathbb Z)$ prepotential \eqref{SL2ZprepPoly} and, for convenience, we have introduced the `$\varGamma_0(N)$ polylogarithms'
\begin{equation}
	\hat{\rm Li}_{k}^{(N)}(z) \equiv {\rm Li}_{k}(z)-\frac{1}{N^k}\,{\rm Li}_{k}(z^N) \,.
\end{equation}
Note that eq. \eqref{HolPrepotCongr2} is cusp-covariant and reduces to the ${\rm SL}(2;\mathbb Z)$ prepotential by setting $N=1$ and $\mathfrak{a}=\infty$, whereby the Kloosterman-Selberg zeta function $ Z_{\infty\infty}(a,b;s)$ simply reduces to its ${\rm SL}(2;\mathbb Z)$ counterpart, $Z(a,b;s)$.

Although the positive frequency modes in $T$ of the generalised holomorphic prepotential do transform covariantly as weight $-2n$ modular forms in the $U$ variable, the zero-frequency mode does not. Clearly, the inhomogeneous polynomial term is completely determined by the transformation properties of the Eichler integral $\tilde{E}_{-2n,\mathfrak{a}}(z)$ under the generators of $\varGamma_0(N)$. Since it is always possible to decompose the latter into repeated applications of the generators $S$ and $T$ of the full modular group, one may uniquely fix the polynomial transformation of $\tilde{E}_{-2n,\mathfrak{a}}(z)$ under any element of $\varGamma_0(N)$ by using eq. \eqref{EichlerCongrFuncRel} and the known transformation properties of the Eichler integral of ${\rm SL}(2;\mathbb Z)$.

Moreover, the behaviour of the generalised holomorphic prepotentials under $\sigma_{T\leftrightarrow U}$ can be straightforwardly determined from eq. \eqref{HolPrepotCongr2} and from the property \eqref{analyticContPoly} of polylogarithms to be
\begin{equation}
	f_n(T,U;\mathfrak{a})-f_n(U,T;\mathfrak{a}) = \delta_{\mathfrak{a}\infty}\,P_{\sigma}(T,U) \,,
\end{equation}
with $T$ and $U$ lying in the fundamental chamber, with real parts in the interval $(0,1)$ and with $P_\sigma(T,U)$ being the polynomial \eqref{Psigma}.


\subsection{Generalised Yukawa couplings for $\varGamma_0(N)$}

In complete analogy to the ${\rm SL}(2;\mathbb Z)$ case, one may compute interactions in the low energy $\mathcal N=2$ supergravity by acting with suitable derivatives on the $\varGamma_0(N)$ generalised holomorphic prepotential. Here we shall restrict our attention to the simplest case of holomorphic generalised Yukawa couplings attached to the cusp $\mathfrak{a}$
\begin{equation}
	Y_{T^{2n+1}}(T,U;\mathfrak{a}) = - 2^{-2n-1}\,D_T^{2n+1}\,f_n(T,U;\mathfrak{a}) = \left(\frac{\partial_T}{2\pi i}\right)^{2n+1}\,f_n(T,U;\mathfrak{a}) \,.
\end{equation}
The degree $2n$ polynomial in the inhomogeneous transformation of $f_n$ is again annihilated by the derivatives and the Yukawa coupling $Y_{T^{2n+1}}$ transforms as a weight $2n+2$ and $-2n$ holomorphic modular form under $\varGamma_0(N)_T$ and $\varGamma_0(N)_U$, respectively. In terms of the Fourier coefficients $F_{\mathfrak{a},n}(M)$ of the Niebur-Poincar\'e series $\mathcal F_{\mathfrak{a}}(1+n,1,-2n)$ expanded around the cusp $\infty$, one finds explicitly
\begin{equation}
	\begin{split}
	Y_{T^{2n+1}}(T,U;\mathfrak{a}) &= 2^{-2n}\sum_{M>0} q_T^M\,\mathcal F_{\mathfrak{a}}(1+n,M,-2n;U) \\
				&= \delta_{\mathfrak{a}\infty}\,Y_{T^{2n+1}}(NT,NU)+2^{-2n} \sum_{M>0\atop L\in\mathbb{Z}} M^{2n+1}\, F_{\mathfrak{a},n}(ML) \mathcal{R}_N(q_T^M\,q_U^L) \,,
	\end{split}	
	\label{YukawasCongr}
\end{equation}
which generalises the ``multi cover'' formula for the standard Yukawa couplings $Y_{T^3}$ to the case of freely-acting $\mathbb Z_N$ orbifolds. To lighten the notation, we have introduced the rational function
\begin{equation}
	\mathcal{R}_N(z) \equiv \frac{z}{1-z}-\frac{z^N}{1-z^N}\,.
\end{equation}
From the above expression, one may readily extract the simple pole divergence at $q_T=q_U$,
\begin{equation}
	Y_{T^{2n+1}}(T,U;\mathfrak{a}) \sim \delta_{\mathfrak{a}\infty}\,\frac{(2n+1)!}{2^{2n}}\,\frac{q_T}{q_U-q_T} \,.
\end{equation}
This singularity, together with holomorphy, modularity and eq. \eqref{YukawasCongr} uniquely fixes the expression of the Yukawa couplings
\begin{equation}
Y_{T^{2n+1}}(T,U;\mathfrak{a}) =2^{-2n} \, \frac{ E_{2n+2,\infty} (T)\,\mathcal{F}_{\mathfrak{a}}(1+n,1,-2n;U) }{j_\infty(T)-j_{\mathfrak{a}}(U)} \,,
\end{equation}
where it is understood that the holomorphic Eisenstein series, the Niebur-Poincar\'e series and the Hauptmodul pertain to the Hecke congruence subgroup $\varGamma _0 (N)$. For $N=1$ one recovers the  ${\rm SL}(2;\mathbb Z)$ result \eqref{YukawasSL2Z}. Moreover, using the results of \cite{Angelantonj:2013eja}, it is straightforward to express the generalised Yukawa couplings $Y_{T^{2n+1}}(T,U;\mathfrak{a})$ in terms of the  holomorphic modular forms of $\varGamma_0(N)$. For instance, one gets
\begin{equation}
\begin{split}
Y_{T^3} (T,U;\infty ) &= \frac{1}{6} \, \frac{ \left[ 4 X^2 (T) - E_4 (T) \right] \left[ E_6 (U) - 2 X (U) E_4 (U) \right]}{\varDelta_8 (U) \, \left[j_\infty (T) - j_\infty (U)\right]} \,,
\\
Y_{T^3} (T,U;0 ) &=\frac{1}{24} \, \frac{ \left[ 4 X^2 (T) - E_4 (T) \right] \left[  E_6 (U) + X (U) E_4 (U) \right]}{\varDelta_8 (U)\, \left[ j_\infty (T) - j_0 (U)\right] } \,, 
\end{split}
\end{equation}
for $N=2$, and
\begin{equation}
\begin{split}
Y_{T^3} (T,U;\infty ) &=\frac{3}{2048} \, \frac{\left[ 9 X^2 (T) - 4 E_4 (T)\right]\, \left[ 9 X^2 (U) - 4 E_4 (U)\right]}{\varDelta_6 (U) \, \left[ j_\infty (T) - j_\infty (U)\right]}\,,
\\
Y_{T^3} (T,U;0 ) &=\frac{3}{2048} \, \frac{\left[ 9 X^2 (T) - 4 E_4 (T)\right]\, \left[  X^2 (U) - 4 E_4 (U)\right]}{\varDelta_6 (U) \, \left[ j_\infty (T) - j_0 (U)\right]}\,,
\end{split}
\end{equation}
for $N=3$. In these expressions $E_w (z)$ are the holomorphic Eisenstein series of ${\rm SL} (2;\mathbb{Z} )$,  $\varDelta_w$ is the weight-$w$ cusp form of $\varGamma_0 (N)$,
and $X (z ) = E_2 (z) - N E_2 (Nz)$ is the weight-two holomorphic modular form of $\varGamma_0 (N)$ \cite{Angelantonj:2013eja}.

The generalisation to higher values of $\kappa$, with $(\kappa,N)=1$, is also straightforward and may be obtained by acting with the Hecke operator on the $\kappa=1$ result,
\begin{equation}
Y_{T^{2n+1}}(T,U,\kappa;\mathfrak{a}) = \kappa^{-n}\, H^{(N)}_\kappa\cdot Y_{T^{2n+1}}(T,U;\mathfrak{a}) \,.
\end{equation}


\section*{Acknowledgements} We are grateful to Per Berglund and Kathrin Bringmann for discussions. C.A. would like to thank the TH Unit at CERN and the  Arnold Sommerfeld Centre at the Ludwig-Maximilians-Universit\"at M\"unchen and the Max-Planck-Institut f\"ur Physik in M\"unchen for hospitality during different stages of this project.
I.F. would like to thank to thank the Physics Department of Torino University for hospitality.
This work was partially supported by the European ERC Advanced Grant no. 226455 ``Supersymmetry, Quantum Gravity and Gauge Fields'' (SUPERFIELDS) and by the  Compagnia di San Paolo contract ``Modern Application in String Theory'' (MAST) TO-Call3-2012-0088.

\appendix


\section{Kloosterman-Selberg zeta function}\label{kloosterman}

The classical Kloosterman sums for the modular group ${\rm SL}(2;\mathbb Z)$ are defined as
\begin{equation}
	S(m,n;c) \equiv\sum_{d\in(\mathbb Z/c\mathbb Z)^\ast}\exp\left[\frac{2\pi i}{c}(m\,d+n\,d^{-1})\right] \,,
\end{equation}
where $m,n$ and $c$ are integers and $d^{-1}$ denotes the arithmetic inverse of $d$ mod $c$. Clearly, it is symmetric under the exchange of $m$ and $n$. One defines the Kloosterman-Selberg zeta function as in \cite{iwaniec2}
\begin{equation}
	Z(m,n;s) \equiv \sum_{c>0}\frac{S(m,n;c)}{c^{2s}}\,.
\end{equation}
In the special case when $mn=0$, the Kloosterman-Selberg zeta function reduces to
\begin{equation}
	Z(0,0;s)=\frac{\zeta(2s-1)}{\zeta(2s)}\ ,\qquad Z(0,m;s)=\frac{\sigma_{1-2s}(m)}{\zeta(2s)} \,,
\end{equation}
where $\sigma_t(m)=\sum_{d|m}d^t$ is the divisor function.

For $mn\neq 0$, it is convenient to introduce the \emph{associated} Kloosterman-Selberg zeta function following  \cite{iwaniec2,Angelantonj:2012gw}
\begin{equation}
\label{KSzetadef}
	\mathcal{Z}(m,n;s) \equiv \frac{1}{2\sqrt{|mn|}}\sum_{c>0}\frac{S(m,n;c)}{c}\times 
			\begin{cases}
				J_{2s-1}\left(\frac{4\pi}{c}\sqrt{mn}\right) & {\rm if} \quad mn>0 \\
				I_{2s-1}\left(\frac{4\pi}{c}\sqrt{-mn}\right) & {\rm if}\quad mn<0 \\
			\end{cases} \,,
\end{equation}
where $I_s(z)$ and $J_s(z)$ are the Bessel $I$ and $J$ functions. Note that $\mathcal Z(m,n;s)$ is related to $Z(m,n;s)$ via \cite{iwaniec2}
\begin{equation}
	\mathcal{Z}(m,n;s) = \pi (4\pi^2|mn|)^{s-1} \sum_{k=0}^\infty \frac{(-4\pi^2 mn)^k}{k!\,\varGamma(2s+k)}\,Z(m,n;s+k) \,.
\end{equation}

The Kloosterman zeta functions $Z$ and $\mathcal Z$ can also be defined in the case of Hecke congruence subgroups of ${\rm SL}(2;\mathbb Z)$. The Kloosterman-Selberg zeta function associated to a pair of cusps $\mathfrak{ab}$ of $\varGamma_0(N)$ is defined for ${\rm Re}(s)>1$ by the absolutely convergent sum
\begin{equation}
	Z_{\mathfrak{ab}}(m,n;s) \equiv \sum_{ \binom{\ a \quad \ast\ }{\ c \quad d\ } \in \varGamma_\infty\backslash \sigma_{\mathfrak{a}}^{-1} \varGamma_0(N) \sigma_{\mathfrak{b}}/\varGamma_{\infty}} \frac{e^{2i\pi(m\frac{d}{c}+n\frac{a}{c})}}{c^{2s}} \,,
	\label{CongrKSZ1}
\end{equation}
of $2\times 2$ real matrices $\binom{\ a \quad \ast\ }{\ c \quad d\ }$ in the double coset $\varGamma_\infty\backslash \sigma_{\mathfrak{a}}^{-1} \varGamma_0(N) \sigma_{\mathfrak{b}}/\varGamma_{\infty}$. Here, $\sigma_{\mathfrak{a}}$ is the scaling matrix \cite{iwaniec2} associated to the cusp $\mathfrak{a}$  and $\varGamma_\infty$ is the stabiliser of the cusp at $\infty$.

Similarly to the ${\rm SL}(2;\mathbb Z)$ case, for $mn=0$, the Kloosterman-Selberg zeta function may be evaluated in terms of the Riemann zeta function. For $N$ prime, one finds
\begin{equation}
	Z_{\infty\infty}(0,0;s)=\frac{N-1}{N^{2s}-1}\,\frac{\zeta(2s-1)}{\zeta(2s)} \ ,\qquad Z_{\infty 0}(0,0;s)=\frac{N^{2s-1}-1}{N^{s-1}(N^{2s}-1)}\,\frac{\zeta(2s-1)}{\zeta(2s)} \,,
\end{equation}
and
\begin{equation}
	\begin{split}
				& Z_{\infty\infty}(0,m;s) = \frac{N\,\sigma_{1-2s}(m/N)-\sigma_{1-2s}(m)}{(N^{2s}-1)\,\zeta(2s)} \,,\\
				& Z_{\infty 0}(0,m;s) = \frac{N^{2s-1}\,\sigma_{1-2s}(m)-\sigma_{1-2s}(m/N)}{N^{s-1}(N^{2s}-1)\,\zeta(2s)} \,,
	\end{split}
\end{equation}
where it is understood that $\sigma_{1-2s}(m/N)$ vanishes unless $N$ divides $m$.

Similarly, for $mn\neq 0$ one defines the \emph{associated} Kloosterman-Selberg zeta function attached to the pair of cusps $\mathfrak{ab}$ via
\begin{equation}
	\mathcal{Z}_{\mathfrak{ab}}(m,n;s) \equiv \frac{1}{2\sqrt{|mn|}}\sum_{ \binom{\ a \quad \ast\ }{\ c \quad d\ } \in \varGamma_\infty\backslash \sigma_{\mathfrak{a}}^{-1} \varGamma_0(N) \sigma_{\mathfrak{b}}/\varGamma_{\infty}} \frac{e^{2i\pi(m\frac{d}{c}+n\frac{a}{c})}}{c} \times 
			\begin{cases}
				J_{2s-1}\left(\frac{4\pi}{c}\sqrt{mn}\right) & {\rm if} \quad mn>0 \\
				I_{2s-1}\left(\frac{4\pi}{c}\sqrt{-mn}\right) & {\rm if}\quad mn<0 \\
			\end{cases} \,.
		\label{CongrKSZ2}
\end{equation}

Notice that the definitions \eqref{CongrKSZ1} and \eqref{CongrKSZ2} valid for any positive integer $N$, include the case of the full modular group for $N=1$, where one recovers the ${\rm SL}(2;\mathbb Z)$ Kloosterman zeta functions
\begin{equation}
	\left. Z_{\infty\infty}(m,n;s)\right|_{N=1} = Z(m,n;s) \ ,\qquad \left. \mathcal{Z}_{\infty\infty}(m,n;s)\right|_{N=1} = \mathcal{Z}(m,n;s) \,.
\end{equation}


\section{Modular properties of the Eichler integral}\label{appEichler}

In this Appendix, following \cite{0990.11041}, we explicitly derive the transformation properties of the Eichler integral $\tilde E_{-2n} (z)$, with even negative weight $w=-2n$, with respect to the action of the modular group ${\rm SL} (2;\mathbb{Z})$. We recall the definition 
\begin{equation}
\tilde E_{-2n} (z) = \frac{\zeta (2n+2)}{2\pi i } \, z^{2n+1} + \frac{\zeta (2n+1)}{2} + \sum_{N>0} \sigma_{-1-2n} (N)\, q^N\,,
\end{equation}
and its general transformation property
\begin{equation}
(c z +d)^{2n} \tilde E_{-2n} \left( \frac{az+b}{cz+d}\right) = \tilde E_{-2n} (z) + P_\gamma (z)\,, \qquad {\rm with}\quad \gamma = \begin{pmatrix} a & b \\ c & d \end{pmatrix}\in {\rm SL} (2;\mathbb{Z})\,,
\end{equation}
and $P_\gamma (z)$ is a degree $2n$ polynomial depending on the transformation $\gamma$ and satisfies the cocycle relation
\begin{equation}
P_{\gamma \gamma '} = P_\gamma \big|_{-2n}\, \gamma ' + P_{\gamma '}\,.
\end{equation}
It is uniquely specified by $P_T (z)$ and $P_S (z)$, with $T$ and $S$ the two generators of the modular group,
\begin{equation}
\label{defTS}
T : \quad z \to z+1 \,, \qquad S : \quad z \to -1/z\,.
\end{equation}

The polynomial $P_T (z)$ can be straightforwardly  derived, and reads
\begin{equation}
P_T (z) =  \frac{\zeta (2n+2)}{2\pi i } \, \left[ (z+1)^{2n+1} - z^{2n+1} \right] =  \frac{\zeta (2n+2)}{2\pi i } \sum_{k=0}^{2n} {2n+1\choose k} \, z^k\,.
\end{equation}

The identification of $P_S (z)$ is a bit more involved. We have by definition
\begin{equation}
z^{2n} \, \tilde E_{-2n} (-1/z) - \tilde E_{-2n} (z) = \sum_{k=0}^{2n} C_k z^k\,,\label{app1}
\end{equation}
for some coefficients $C_k$. Note that the simple change of variable $z \to -1/z$ implies the relations 
\begin{equation}
C_k = (-1)^{k+1} \, C_{2n-k} \qquad {\rm for}\quad k=0,1,\ldots , 2n\,,
\end{equation}
and, a result, only half of the coefficients are independent. To determine the independent coefficients let us consider the case where $z=iy$, and define the function
\begin{equation}
\begin{split}
\tilde e_{-2n} (y) &= \sum_{N>0} \sigma_{-1-2n} (N) \, e^{-2\pi N y}
\\
&= \tilde E_{-2n} (iy) - \frac{\zeta (2n+1)}{2} - \frac{\zeta (2n+2)}{2\pi}\, (-1)^n \, y^{2n+1}\,.
\end{split}
\end{equation}
From eq. \eqref{app1} one finds 
\begin{equation}
\tilde e_{-2n} (y) - (-1)^n y^{2n}\,  \tilde e_{-2n} (1/y)  = \sum_{k=-1}^{2n+1} A_k \, y^k\,,
\label{app2}
\end{equation}
where
\begin{equation}
\begin{split}
A_{-1} &= \frac{\zeta (2n+2)}{2\pi}
\\
A_{2n+1} &= (-1)^{n+1} \frac{\zeta (2n+2)}{2\pi}\,, 
\end{split}\qquad
\begin{split}
A_0 &= - C_0 -\frac{ \zeta (2n+1)}{2}\,,
\\
A_{2n} &= (-1)^n \frac{ \zeta (2n+1)}{2} - (-1)^n C_{2n}\,,
\end{split}
\end{equation}
and $A_m = - i^m C_m$, for $m=1,\ldots, 2n-1$.

The coefficients $A_k$, and thus the $C_k$, can then be determined by computing the Mellin transform
\begin{equation}
L^\star (s) = \int_0^\infty d y \, y^{s-1}\, \tilde e_{-2n} (y)\,.
\end{equation}
A straightforward evaluation of the integral yields
\begin{equation}
L^\star (s) = (2\pi)^{-s} \, \varGamma (s) \, \zeta (s) \, \zeta (s +2n+1)\,.
\label{app3}
\end{equation}
Alternatively, using the relation \eqref{app2}, we can write
\begin{equation}
L^\star (s) = \int_1^\infty dy \, y^{s-1} \, \tilde e_{-2n} (y) + (-1)^n\int_0^1 dy \, y^{2n+s-1}\, \tilde e_{-2n} (1/y ) + \sum_{k=-1}^{2n+1} \frac{A_k}{s+k}\,.
\label{app4}
\end{equation}
By matching the singularities of the two expressions \eqref{app3} and \eqref{app4} one thus finds the following expression for the inhomogeneous contribution
\begin{equation}
P_S (z) = - \frac{(2\pi i )^{2n+1}}{2} \sum_{k=1}^n \frac{B_{2k}\, B_{2 n - 2k +2}}{(2k)!\, (2n-2k+2)!}\,  z^{2k-1}\,,
\end{equation}
that generalises \cite{0990.11041} to the cases $n>1$.

Alternatively, the polynomials $P_\gamma$ could be computed using the method of \cite{zbMATH06149482}.



\providecommand{\href}[2]{#2}\begingroup\raggedright\endgroup

\end{document}